\documentclass[traditabstract]{aa} 
\usepackage{graphicx}
\usepackage{longtable,lscape}
\usepackage{txfonts}
\usepackage{rotating}

\usepackage{latexsym}
\usepackage{psfig}
\usepackage{epsfig}

\begin{document}

\newcommand{\be}{\begin{equation}}
\newcommand{\ee}{\end{equation}}
\newcommand{\cmq}{cm$^{-3}$}
\newcommand{\um}{$\mu m$}
\newcommand{\sori}{$\sigma$~Ori}
\newcommand{\roph}{$\rho$-Oph}
\newcommand{\Msun}{M$_\odot$}
\newcommand{\Lsun}{L$_\odot$}
\newcommand{\Ha}{H$\alpha$}
\newcommand{\Pab}{Pa$\beta$}
\newcommand{\Pag}{Pa$\gamma$}
\newcommand{\Brg}{Br$\gamma$}
\newcommand{\Lpag}{L(Pa$\gamma$)}
\newcommand{\Teff}{T$_{eff}$}
\newcommand{\Lstar}{$L_{*}$}
\newcommand{\Rstar}{$R_{*}$}
\newcommand{\Mstar}{$M_{*}$}
\newcommand{\Lacc}{L$_{acc}$}
\newcommand{\Macc}{$\dot M_{acc}$}
\newcommand{\Mloss}{$\dot M_{loss}$}
\newcommand{\Mwind}{$\dot M_{wind}$}
\newcommand{\Ldisk}{L$_{disk}$}
\newcommand{\Myr}{M$_\odot$/yr}

\newcommand{\simless}{\mathbin{\lower 3pt\hbox {$\rlap{\raise 5pt\hbox{$\char'074$}}\mathchar"7218$}}} 
\newcommand{\simgreat}{\mathbin{\lower 3pt\hbox
     {$\rlap{\raise 5pt\hbox{$\char'076$}}\mathchar"7218$}}} 


\title{U-band study of the accretion properties in the \sori\ star-forming region
\thanks{
Based on observations collected at the European Southern Observatory, Chile.
Program 082.C-0313(A).}
}

\author{
E. Rigliaco\inst{1,2},
A. Natta\inst{1},
S. Randich \inst{1},
L. Testi\inst{1,3},
\and
K. Biazzo\inst{1}
}

\institute{
    Osservatorio Astrofisico di Arcetri, INAF, Largo E.Fermi 5,
    I-50125 Firenze, Italy
\and
Universit\`a  di Firenze, Dipartimento di Astronomia, Largo E.Fermi 2,
    I-50125 Firenze, Italy
\and
ESO, Karl-Schwarschild Strasse 2, D-85748 Garching bei M\"unchen, Germany
}

\offprints{erigliaco@arcetri.astro.it}
\date{Received 29 July 2010; accepted 9 October 2010}

\authorrunning{Rigliaco et al.}
\titlerunning{Accretion in $\sigma$Ori}

\abstract
{This paper presents the results of an U band survey with FORS1/VLT
of a large area in  
the \sori\ star-forming region.
We combine the U-band photometry with
literature data to compute accretion luminosity and mass accretion rates
from the U-band excess emission for all objects (187)
detected by
{\it{Spitzer}} in the FORS1 field  and classified by
Hernandez et al. (2007)  as likely members of the cluster.
The sample stars range in mass from $\sim 0.06$ to $\sim 1.2$ \Msun;
72 of them show evidence of disks and we measure mass accretion rates
\Macc\ between $<10^{-11}$ and few $10^{-9}$ \Msun/y, using
the colors of the diskless stars as photospheric templates.
Our results confirm the
dependence of \Macc\ on the mass of the central object,
which is stronger for low-mass stars and flattens out for masses
larger than $\sim 0.3$ \Msun; the spread of \Macc\ for any value of the stellar mass is $\sim$2 orders of magnitude. We discuss the implications of these results in the context of disk evolution models.
Finally, we analyze the relation between \Macc\ and the excess emission in the {\it{Spitzer}} bands, 
and find that at \Macc\ $\sim 10^{-10}$ \Msun/y the inner disks change from optically thin to optically thick.
}


\maketitle

\section {Introduction}

In recent years, our knowledge of the properties of young stars in several star-forming regions has made enormous progress.
In particular,
{\it{Spitzer}} observations have provided new information on the IR properties
of circumstellar disks,
and the discussion on how disks evolve in time has  gained new 
momentum. In addition to the classical viscous evolution, which
dissipates disks by accreting their matter onto the
central stars, other processes such as gravitational
instabilities, photoevaporation by  X-ray and UV
radiation of the central star (Hollenbach et al.~2000; Gorti \& Hollenbach~2009),  and planet formation (Dullemond et al.~2007)
 have been recognized to be important. Photoevaporation and planet
formation may  both  shape the SED (spectral energy distribution) of
so-called transitional disks, which 
have very low emission in the near- and mid-IR and strong excess emission
at longer wavelengths 
(Calvet et al.~2002,~2005; D'alessio et al.~2006; Currie et al.~2009).
In all cases, disks seem to come in all variesties:
it is clear that neither time nor the mass of the system
control them uniquely, and it seems very likely
that disk properties and evolution
depend also on
the initial conditions, i.e., on the properties of the
molecular core from which the star+disk system forms (Hartmann et al.~1998,~2006; Dullemond et al.~2006, 
Clarke~2007, Vorobyov \&\ Basu~2009).

An important contribution to this discussion  comes from
measurements of the mass accretion rate for well characterized
samples of stars.  
However, there are only few systematic determinations of the mass-accretion rates
in large samples of objects within the same star-forming regions,
covering a large range of central masses,
with well measured SEDs and complete to include also
diskless stars, limited so
far  to  \roph\ (Natta et al. 2006) and  Tr 37
(Sicilia-Aguilar et al. 2010).

In this paper, we add a third region, \sori, to the list.
The \sori\ cluster is ideally suited for this kind of study. It contains
more than 300 young stars, ranging in mass from the bright, massive multiple system 
\sori\ itself (the spectral type of the brightest star is O9.5V, Caballero.~2007) to brown dwarfs. 
It is located at 
a distance of $\sim$ 360 pc ({\it{Hipparcos}} distance $352^{+166}_{-85}$ pc for the O9.5V star
Brown et al.~1994; Perryman et al.~1997) 
and has an age of $\sim$ 3Myr (Zapatero-Osorio et al.~2002; Oliveira et al.~2004). 
The region has negligible extinction (Bejar et al.~1999;
Oliveira et al.~2004), and
has been extensively studied  in the optical, X-ray and 
infrared (e.g., Kenyon et al.~2005; Zapatero-Osorio et al.~2002; 
Jeffries et al.~2006;
Franciosini et al.~2006; Hernandez et al.~2007; Caballero et al.~2007; Wolk~1996). 
Hernandez et al.~(2007) have obtained  {\it{Spitzer}} images of a large
area in \sori\ in the four IRAC bands and with MIPS at 24 $\mu$m;
they find 336 candidate members,
of which 66\% are class III stars and 34\% show evidence of disks.


Accretion rates  have been obtained
by Gatti et al.~(2008) for 35  objects in \sori\
from the luminosity of the
near-IR hydrogen line Pa$\gamma$; they found mass accretion rates lower 
on average than in younger regions. However, their sample was small and 
limited in mass (0.12--0.5 \Msun). 
In this paper, we present the results for a much larger and better
characterized sample from Hernandez et al.~(2007). 
We measure mass accretion rates from the U-band excess emission, which originates
in the  accretion shock where accreting matter impacts on the stellar surface
(Gullbring et al.~1998; Calvet \& Gullbring~1998). The correlation between
the U-band excess luminosity and the accretion luminosity has been
established both empirically (Gullbring et al.~1998; Herczeg \&\ Hillenbrand~2008)
and theoretically (Calvet \&\ Gullbring~1998).
The U-band excess is an excellent proxy of the accretion luminosity, which
allows obtaining reliable values of the mass accretion rate for large
samples of stars using  little observing time when, as
in \sori, the extinction is negligible.
It very well complements measurements obtained from other tracers, such as
the IR hydrogen recombination line luminosities (see the discussion in Herczeg
\&\ Hillenbrand~2008).

The paper is organized as follows:
observations and data reduction are described in Sect.~\ref{obs}, 
the properties of the sample are derived in Sect.~\ref{sample}, 
in  Sect.~\ref{method} we discuss
the method used to derive the accretion properties. 
The results are discussed in Sect.~\ref{result} and \ref{discussion}. 
Three appendices present additional material on: 
the recomputation of accretion rates in \roph\ with the new distance and evolutionary tracks; 
the accretion properties of BD candidates beyond the Spitzer sample; 
and the properties of transitional and evolved disks in our sample.

\section {Observations and data analysis}
\label{obs}

\subsection {U-band photometry}

\begin{figure}[htbp]
\centering
\includegraphics[width=9cm]{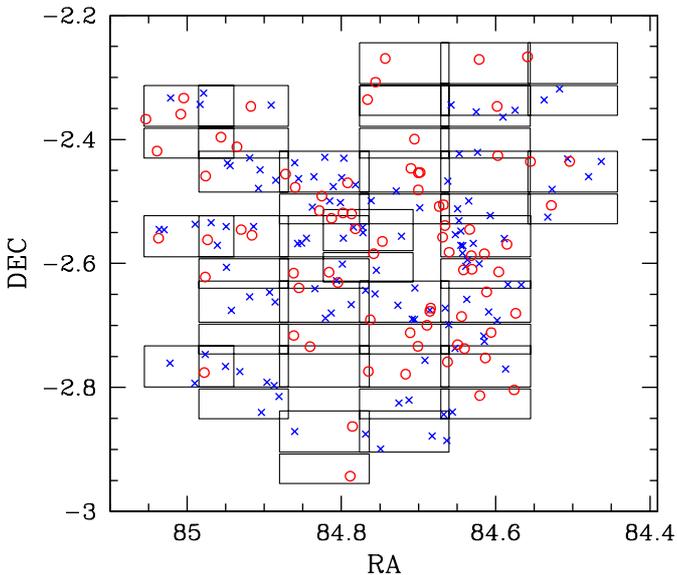} 
\caption{Fields in the \sori\ cluster that have been imaged with FORS1 (solid boxes).
Crosses show class III stars (blue in the online version of the figure), 
circles class II, TD and EV object (red in the online version of the figure), 
as classified by Hernandez et al.~(2007). 
}
\label {CV}
\end{figure}

We have performed a U-band survey that covers a total field of 
$\sim$1000 arcmin$^2$ in the $\sigma$~Orionis cluster.
Observations were carried out with 
FORS1 mounted on the UT2 telescope at the VLT using $\rm{u\_HIGH}$ filter and 
were performed in service mode 
during seven nights from October 2008 to March 2009. 
All the nights were photometric, with seeing in the range 0.6-1.5". 
We observed 28 fields (FOV 6'.8 $\times$ 6'.8) in \sori\ 
with an exposure time of 900 seconds each. 
The distribution of the pointings obtained with 
FORS1 is shown in Figure~\ref{CV} and the log of the observation is provided in 
Table~\ref{chronoList}. 
The long exposure time on each field allowed us to reach a U-band 
limiting magnitude of $\sim$23;
objects brighter than U$\sim$17 mag are saturated.
The fields around
the two brightest stars of the quintuplet system containing 
$\sigma$~Ori 
(within 0.2 pc projected distance) 
had to be excluded because of to light contamination.

\begin{table}
\centering
\caption{Journal of observations. RA and DEC refer to the pointing center.}
\begin{tabular}{c c c c c c}
\hline \hline
 Run dates & Observed field & RA & DEC & Seeing\\
 (y-m-d) &  & ($\circ$) & ($\circ$) &  (") \\
\hline
2008-10-05 & field 1 & 84.5001 & -2.30238 & 0.92 \\
 & field 3 & 84.7184 & -2.30234 & 1.33\\
2008-10-06 & field 6 & 84.9268 & -2.37146  & 1.20\\
 & field 7 & 84.9976 & -2.37146  & 1.32\\
 & field 8 & 84.5002 & -2.47742 & 1.51\\
 & field 10 & 84.7184 & -2.47742 & 1.46\\
 & field 11 & 84.8222 & -2.47742 & 1.19\\ 
2008-10-08 & field 13 & 84.6129 & -2.58167  & 0.65\\
 & field 15 & 84.8222 & -2.58167  & 0.96\\
 & field 16 & 84.9268 & -2.58167  & 0.95\\
 & filed 18 & 84.613 & -2.68734 & 1.0\\
 & field 19 & 84.7185 & -2.68734  & 0.89\\
 & field 20 & 84.8222 & -2.68734 & 0.99\\
2008-12-21 & field 2 & 84.613 & -2.30234  & 0.81\\
 & field 4 & 84.613 & -2.37146 & 0.76\\
 & field 5 & 84.7184 & -2.37146 & 0.72\\
 & field 9 & 84.6131 & -2.47742 & 0.79\\
 & field 22 & 84.613 & -2.79169 & 0.74\\
 & field 23 & 84.7184 & -2.79169  & 1.09\\
 & field 27 & 84.7184 & -2.87086  & 0.73\\
 & field 28 & 84.8221 & -2.89632 & 0.68\\
2009-01-24 & field 12 & 84.9267 & -2.45196  & 0.95\\
 & field  21 & 84.9267 & -2.68734 & 0.96\\
 & field 24 & 84.8221 & -2.76621 & 1.06\\
2009-01-30 & field 17 & 84.9976 & -2.55619  & 0.99\\
 & field 26 & 84.9974 & -2.76621 & 0.78\\
2009-03-25 & field 14 & 84.7653 & -2.57167  & 0.70\\
 & field 25 & 84.9268 & -2.79169 & 0.69 \\
\hline
\end{tabular}
\label{chronoList}
\end{table}

\subsection {Data reduction}
\label{data reduction and error}
The data were reduced using standard procedures including bias subtraction and flat fielding 
within the IRAF \footnote{{\sc iraf} is distributed by National Optical Astronomy
Observatories, which are operated by the Association of Universities
for Research in Astronomy, Inc., under cooperative agreement with the
National Science Foundation.} package.
We performed aperture photometry with the PHOT task in the APPHOT package, 
and using noao.digiphot.daophot routines for the photometry extraction. 
The stellar density was generally low enough to make aperture photometry acceptable. 
The IRAF routine MKAPFILE was used to determine and apply aperture corrections 
based on ensemble averages of stars in each separate frame. 
Astrometry correction was done to center the telescope coordinates.  
Photometric standard stars from Landolt~(1992) and Persson et al.~(1998) were observed 
at least once during each night and were used to flux-calibrate the images using the task PHOTCAL 
and to set the zero point magnitudes of each observing night and for both  chips.
Aperture photometry was performed using ten different apertures per image 
(0.5, 0.6, 0.7, 0.8, 0.9, 1.0, 1.25, 1.5, 2.0, 3.0 times the average FWHM). 
The inner radius of the sky annulus, which allowed us to define the sky 
brightness, was 10 times the average FWHM, 
while the width of the annulus was fixed at 10 pixels. 

The uncertainties on the U-band magnitudes obtained are on the order of $\pm 0.1$ mag.
They are dominated by systematic errors, the biggest of which is the error on the zero point magnitude (about 0.08 mag), which
affects all measurements in the same manner; a second systematic term, on the order of 
0.01 mag, is due to
the color term correction 
with respect to the $\rm{u\_HIGH}$ filter. 
Random errors owing to the aperture photometry technique 
used to derive the stellar flux and the sky brightness are also very small.\\

\subsection {U-band variability}

Young pre-main sequence stars are known to be variable, with timescale
from hours to several days (e.g., Gomez de Castro et al.~1998; Hillenbrand et al.~1998; Brice\~{n}o et al.~2001; Sicilia-Aguilar et al.~2005a, 2005b). 
The U-band variability is probably related to
variations of the accretion rate, and, although a proper study is well outside
the scope of this paper, it is interesting to estimate how large an effect this
is likely to be.

In our FORS1 data there are 30 objects that lie at the superposition of two different fields 
and have therefore been observed twice, with separations that range between few days and few months.
Of these, 19 are Class III stars and show no variability within the photometric
uncertainty. Of the 11 Class II objects, 5 show no variability, 5 have
variations between 0.15 and 0.4 mag, and one (SO866) has two measurements
which differ by about 0.8 mag. 
This is similar to what is observed in other star-forming regions (see, e.g.,
Sicilia-Aguilar et al.~2010 and references therein).

\begin{figure}[htbp]
\centering
\includegraphics[width=9cm]{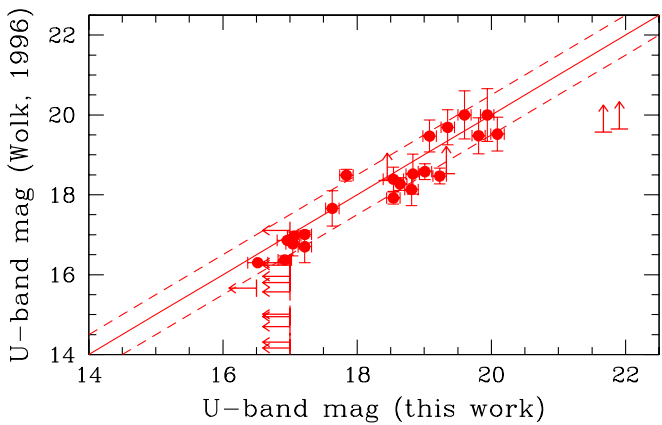} 
\caption{U-band magnitude from Wolk (1966) and
this work. Dashed lines correspond to an interval of $\pm 0.5$ mag.
}
\label{WolkUband}
\end{figure}

Variability on a much longer time base can be checked by comparing our sample
to that observed
by Wolk~(1996), where the observations were carried out in January 1996 
with the 1 meter telescope CTIO. 
The two samples have $\sim$30 class II stars in common 
(see Fig.~\ref{WolkUband}). 
The comparison between the U band magnitude determinations 
shows a $\Delta$U variation of at most 0.5 magnitudes, not different from what
we observe on a shorter timescale. 

A difference of 0.5 mag in the measured U-band magnitude
corresponds to a \Lacc\ and \Macc\ difference  of a factor of at most two (see \S 4~). 
This can be important when discussing individual objects. However, if the
accretion properties of a large sample of stars are considered, it can only
cause a moderate spread in the accretion values. 

\subsection {Spectroscopy}

Among the objects observed with FORS1, six have also optical spectroscopy 
obtained with SARG@TNG.
We obtained spectroscopic observations in 2009 at the {\it Telescopio
Nazionale Galileo} 
(TNG) during three
nights from 27 to 29 January. 
The SARG spectrograph attached to the 3.58m telescope was used with the
2048$\times$4096 CCD 
detector (pixel size = 13.5\,$\mu$m) and the Yellow Grism CD\#3 as
cross-disperser. This allowed 
us to cover $\sim$6200-8000 \AA~wavelength range. We used 
the slit \#1 obtaining $R=$29\,000 as spectral resolution.

The data reduction was performed by using the {\sc echelle} task of the
IRAF package following the 
standard steps of background subtraction, division by a flat-field
spectrum given by a halogen lamp, wavelength 
calibration using emission lines of a Thorium-Argon lamp, and sky
subtraction. 

With exposure times of 10-30 minutes we achieved signal-to-noise
ratios ($S/N$) in 
the range 10--15 in the lithium line $\lambda$6708 region, depending on
airmass and sky conditions. 
Each star was observed 1-3 times. All the spectra acquired per star were
shifted in wavelength 
for the heliocentric correction and then co-added obtaining a $S/N$
ratio in the continuum around 20.

\section {The observed sample} 
\label{sample}

Table~\ref{phot} gives the measured U-band magnitude of all the objects in the 
FORS fields listed as \sori\ members by
Hernandez et al.~(2007) based on optical and near-IR
photometry. 
This sample of 187 objects (out of 336 members) is our basic sample and will
be discussed below.
 
The Hernandez et al.~(2007) sample spans the mass range from $\sim 0.06$ to
2--3 \Msun, and is practically complete above 0.1 \Msun.
Based on the SED in the 
IRAC spectral range (from 3.6 \um\ to 8.0 \um\ ) the {\it Spitzer} sources are
divided in
in 
class II stars, pre-main sequence stars with IR excess typical of optically thick disks 
(classical TTauri stars CTTs or CII), class III stars, with typical colors of stellar photospheres 
(weak-line TTauri stars WTTs or CIII), and stars with non-classical disks,
in turns divided in "evolved disks" (EV), with small excess emission at all 
infrared wavelengths,
and "transitional disks" (TD), which have zero or very low emission
in the near infrared but normal excess at longer wavelengths.
Below we will refer to class II, TD, and EV objects as "disk objects".
 
Our sample includes 115 class III members,
and 72 
objects with evidence of disks.
Among these, there are  54 class II stars, 4 TD (out of the 7 possibly identified by
Hernandez et al. 2007), and 14 EV disks.
Of the 72  stars with disks, 54 are detected in the U-band, 6 are non-detections, 
and 12 are saturated.
Out of the 115 class III stars, 83 are detected, 7 are not, and
25 are saturated. 

The spatial distribution of the observed sample is shown in Fig.~\ref{CV}.

\subsection {Stellar properties}
\label{sectionHR}

Spectroscopically determined spectral types exist for a small fraction of the 
\sori\ objects only 
(Zapatero-Osorio et al.~2002; Barrado y Navascues et al.~2003; 
Muzerolle et al.~2003). 
Therefore, we determine the stellar parameters (effective temperature, luminosity, mass, and radius) 
of all  objects in a homogeneous way  from multi-color photometry. \\

Table~\ref{phot} reports  for each object  broad-band magnitudes collected
from the literature.
The optical photometry is from 
Sherry et al.~(2004); Zapatero-Osorio et al.~(2002); Kenyon et al.~(2005); B\'ejar et al.~(2001) 
and Wolk~(1996).   
The JHK magnitudes are taken from the Two Micron All Sky Survey (2MASS) (Cutri et al.~2003). 
The magnitudes in the four channels of the Infrared Array Camera (IRAC; 3.8-8.0 $\mu$m) 
and the first channel of the Multiband Imaging Photometer for {\it{Spitzer}} (MIPS; 24 $\mu$m) 
are  from Hernandez et al.~(2007). 
When two or more magnitude determinations for the same band were available, 
we choose, if possible, measurements obtained by the same author.

We derived 
the effective temperatures  of each star by comparing the observed magnitudes
to the synthetic colors computed from the model atmosphere of Baraffe et al.
(1998) for $\log g$=4.0 and the appropriate filter passband and zero fluxes.
Luminosities were then computed using the I-band magnitude and the bolometric correction for ZAMS
stars and the \Teff\ -- spectral type correlation of Kenyon \&\ Hartmann
(1995) and Luhman et al.~(2003).
We assumed a distance of D=360 pc, and negligible extinction in all bands (Brown et al.~1994; Bejar et al.~1999). 

We performed two checks on the \Teff\ estimates. In a first test, we compared the
values derived from the model atmosphere synthetic colors
with those obtained by comparing the observed colors (V-R), (V-I) and (R-I)
to those of  ZAMS stars. The differences
are  within 150 K for 90\% of the stars, with a slight systematic
tendency toward higher values for model atmosphere estimates. 
A second test is given by the comparison of our estimates of \Teff\ with 
the spectroscopic determination  derived 
from low- and medium-resolution  observations. In a sample of
22 objects in the range $\sim$ 3200--4000 K we find differences 
of $\pm 150$ K at most.

\begin{figure}[t]
\centering
\includegraphics[width=9.5cm]{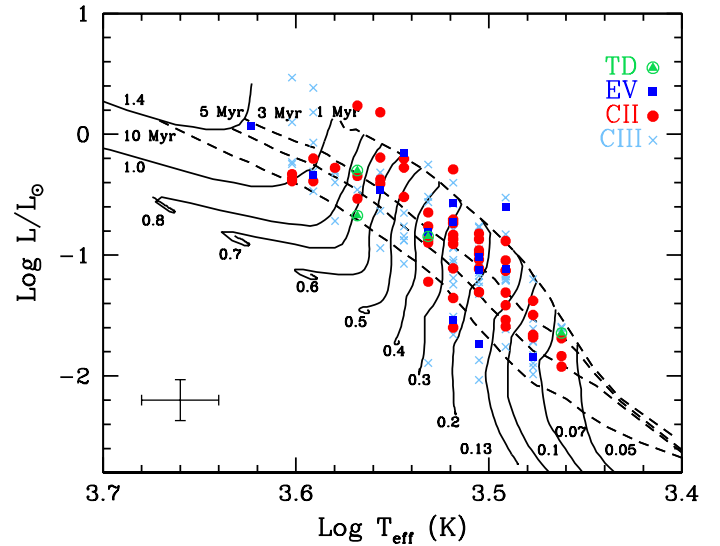} 
\caption{ Location of the observed objects
(class III, class II, and objects with "non-classical" disks,
as labeled) on the HR diagram. 
Evolutionary tracks and isochrones are taken from Baraffe et al.~(1998). Stellar masses and
ages are labeled. 
The horizontal bar refers to the error on the effective temperature, 
the vertical bar 
reflects the error of the bolometric correction of the ZAMS stars related to the error on the 
effective temperatures.
}
\label {HRbar}
\end{figure}

The distribution of the stars in the HR diagram is shown in Fig.~\ref{HRbar} with the 
evolutionary tracks and isochrones from Baraffe et al.~(1998).
The sample covers the mass range between $\sim$1.2 to $\sim$0.05 \Msun;
both the lower and the higher mass limit reflect the sensitivity limit
of the {\it Spitzer} survey (see discussion in Hernandez et al.~2007).
The median age is about 3 My, with a rather large spread,
which is similar for Class II and Class III objects. This spread also remains when only
radial velocity confirmed members  are considered
(Sacco et al.~2008; Kenyon et al.~2005). 
The issue of the spread in age in \sori\ as well as in other young star-forming regions 
has been discussed in several papers (e.g. Hillenbrand~2009 and references therein) 
and further discussion is beyond the purpose of this paper.
Note, however, that the error bars
may be quite large and affect the age estimates significantly.

The stellar parameters of all Class II objects are summarized in Table~\ref{paramCII}.

\section { Mass accretion rate }
\label{method}

Matter accreting from the disk onto the star, channeled along field lines,
shocks at the stellar surface.
About half of the accretion luminosity is released with a typical color temperature
of $\sim 10^4$ K, i.e., much hotter than the stellar photosphere
(Hartigan et al.,~1991, Gullbring \& Calvet,~1998).
The resulting excess emission is clearly detected at short wavelengths, 
in the U-band in particular. It has been shown
(Gullbring \& Calvet~1998, Herczeg \& Hillendrand~ 2008)
 that the U-band excess luminosity is an accurate proxy of the accretion
luminosity, which can be reliably used to measure $L_{acc}$ for T Tauri stars
and BDs.

\subsection {Class III sources}

In order to measure the excess 
of luminosity in U-band for a given object, we need to know 
the measured U-band luminosity and the expected photospheric U band luminosity 
for a non-accreting star with the same parameters.
This is not trivial,
as young stars tend to have a significant level of chromospheric activity
that causes continuum emission at short wavelengths; this 
is not related to accretion and  must, therefore, be counted as
"photospheric" contribution for our purpose.

\begin{figure}[htbp]
\includegraphics[width=9cm]{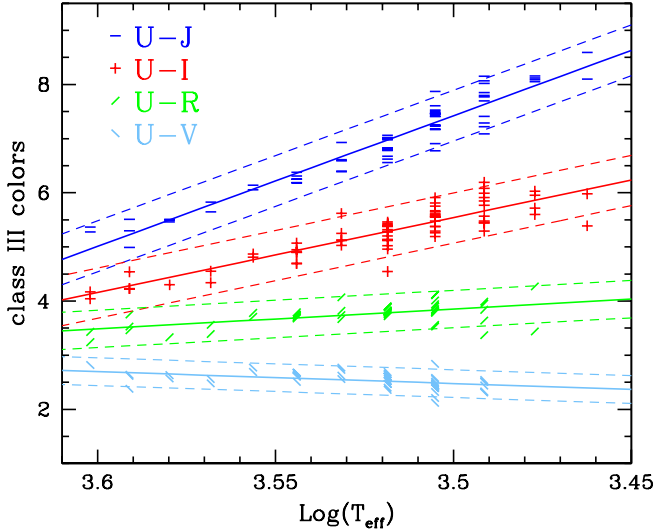}%
\caption{Colors of Class III stars vs. effective
temperature.  Blue horizontal lines plot (U-J), red crosses (U-I),
green 45deg lines (U-R), cyan -45deg lines (U-V), as labeled.
The solid lines show the best fit, 
the dashed lines  $\pm 2\sigma$. In the text, we will define the region between
the dashed lines as the photospheric strip. 
(A color version of this figure is available in the online journal.)}
\label {calib}
\end{figure}

We take advantage of the large number of Class III stars in our sample to
define the typical 
colors of non-accreting young stars.
Fig.~\ref{calib} plots the class III  
(U-J), (U-I), (U-R) and (U-V) colors 
as a function of the effective temperature of the star. 
For each color index, there is a tight correlation with
the effective temperature; the solid lines show the best fits, the dashed lines
the 2$\sigma$ errors, which are $\sim 0.5$ mag at most. 
Hereafter, we will refer to the $\pm 2\sigma$ as the photospheric strip.
Note that while (U-J) increases with $T_{eff}$, (U-V) slightly decreases.

\subsection {Disk sources}

To derive the U-band excess emission in class II objects we assign to each of them
the  photospheric colors of class III stars of the same \Teff,
according to 
the correlations shown in Fig.~\ref{calib}; we  assume that
there is negligible excess emission in the V, R, I, and J band
and derive the U-band excess from the difference between
the observed and the photospheric colors.
We  use (U-I), as I-band magnitudes are available
for all stars in our sample (see Table~\ref{phot}), using the relation:

\begin{equation}
\rm{\Delta U_{excess} = (U-I)_{obs} - (U-I)_{phot} ,}
\label{exc}
\end{equation}
where $\rm{(U-I)_{obs}}$ are the observed $\rm{(U-I)}$ color, 
and $\rm{(U-I)_{phot}}$ is the assigned photospheric color.

We define as accreting all stars with $\rm{\Delta U_{excess}}$ larger than 
the 2$\sigma$ uncertainties of the photospheric colors, as derived in \S 4.1. 
Class II stars with colors 
within the class III photospheric strip will be considered not-accretors
and we can only assign upper limits to the 
accretion luminosity and mass accretion rate.
We choose the 2$\sigma$ uncertainty as good compromise 
 not to loose low-accreting stars.
After defining the accreting or non-accreting stars, we computed 
the excess flux in the U-band as:
\begin{equation}
F_{U,exc}=F_{0,U} \times \big(10^{-U_{obs}/2.5}-10^{-{((U-I)}_{phot}+I_{obs})/2.5}\big) , 
\label{Fuexc}
\end{equation}
where $F_{0,U}$ is the zero point flux in the U-band.

\begin{figure}[htbp]
\includegraphics[width=9cm]{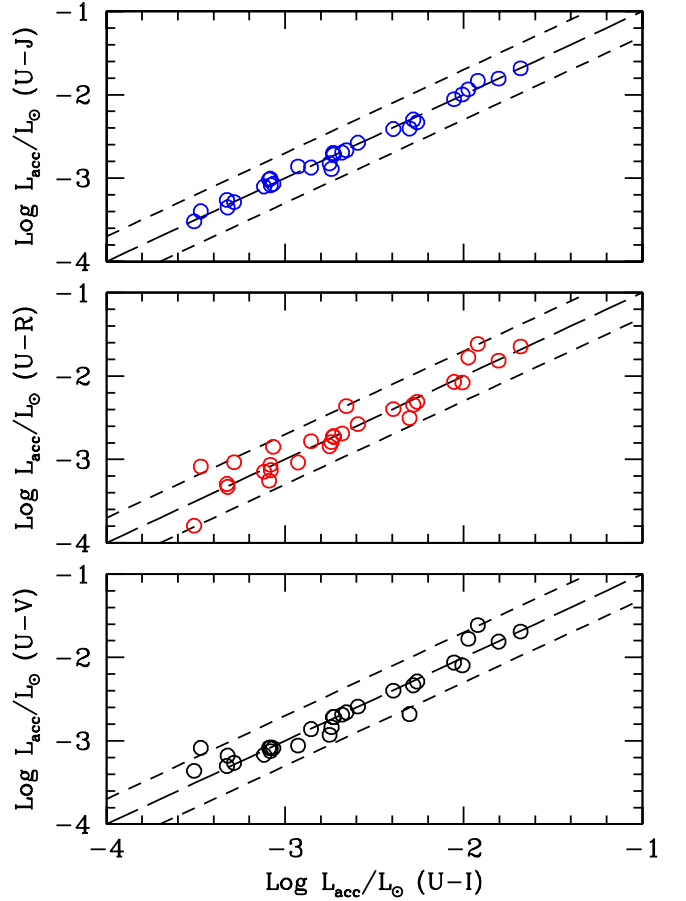}%
\caption{Values of the accretion luminosity
derived from (U-J), (U-R), (U-V) vs. the value derived from (U-I),
as described in the text for a sub-sample of 30 objects.
The dashed lines correspond to $\pm 0.5$ in  $\log \rm{L_{acc}}$.
}
\label{LaccLacc}
\end{figure}

Figure~\ref{LaccLacc} shows the comparison between the accretion luminosities computed 
from  (U-I) vs those derived from (U-J), (U-R) and (U-V), respectively,
for all objects with available photometry. 
The derived \Lacc\ are the same within a factor $\sim 3$ for all objects, and for
70\% of them the agreement is within a factor 1.5.
These results support several aspects of our procedure: 
within the above uncertainties: firstly, our assumption that reddening is negligible at all
wavelengths; secondly, that there is negligible excess emission in 
V, R, I, and J; thirdly, since the photometric data are
collected from the literature and are not simultaneous, that variability
is not the major limiting factor in deriving the accretion properties 
(although it may introduce some scatter in the measurements).\\
We use the (U-I) 
color to define the accretion rates. Assuming an uncertainties in U-band 
of 0.5 mag we estimate an error on the U-band luminosities of a factor 3 at most.

\subsection {Accretion rate}
\label {ACCRrate}

The U-band luminosity obtained from the excess flux in the U-band (Eq.~\ref{Fuexc}) 
is converted into total 
accretion luminosity \Lacc, which is roughly the amount of energy released 
by gas that accretes onto the star, using 
the approximately linear relation derived for  T Tauri stars 
(Gullbring \& Calvet~1998) and brown dwarfs  
(Calvet \& Gullbring~1998; Herczeg \& Hillendrand~2008): 
\begin{equation}
\log \big({{L_{acc}} \over {L_\odot}}\big) = \log \big({{L_{U}} \over {L_\odot}}\big) + 1.0 .
\label{eqLacc}
\end{equation}

Knowing $M_{*}$, $R_{*}$ and the 
accretion luminosity \Lacc\,  we derive the mass accretion rate ${\dot M_{acc}}$ with  
the relation 
\begin{equation}
{\dot M_{acc}} = \big(1 - {{R_{*}} \over {R_{in}}}\big)^{-1}  {{L_{acc} \, R_{*}}\over{G \, M_{*}}} \sim 1.25 {{L_{acc} \, R_{*}}\over{G \, M_{*}}} ,
\end{equation}
where G is the universal gravitational constant and 
the factor $(1-{{R_{*}} \over {R_{in}}})^{-1}$ $\sim$ 1.25 is estimated by assuming 
that the accretion gas falls onto the star from the truncation radius of the disk 
($R_{in}$ $\sim$5$R_{*}$; Gullbring et al.~1998). \\

Table~\ref{paramCII} reports 72 actual values of \Macc. Thirty stars have \Macc\ detection.  
In 30 cases, we can only
estimate upper limits to \Macc. Of these, 6 are objects with U-band emission
below our detection limit, while 24 have colors within the photospheric strip
(\S 4.2). Eight stars have expected U-band photospheric emission brighter
than our saturation limit (labeled~"sat"), and 4 stars have U $< 17$ mag, but colors
with a lower limit than \Macc\ can be safely estimated 
because their photospheric contribution is lower than the saturation limit, 
and the upper limit in the U-band can be only due to the accretion process.

The lowest values of \Macc\ we can estimate range from $\sim 10^{-11}$
\Msun/y for very low, cold objects to few $10^{-10}$ \Msun/y for solar-mass
stars. This trend occurs because
the minimum detectable value of \Macc\ from  U-band photometry
depends not only on the depth of the photometry, but also on the 
physical properties of the star, and on how well one can
estimate the photospheric flux. Indeed, this is at present
the major limiting factor. 

The uncertainties on individual measurements of
\Lacc\ and \Macc\ are quite large.
They come from the combination of  photometric errors and variability, 
the definition of the class III colors,
the adopted relation between the U-band excess luminosity and the accretion
luminosity, and between the latter and \Macc\ and the uncertainty
on the value of \Rstar/\Mstar (see also Sicilia-Aguilar et al.~2010).
Some of these uncertainties have been discussed in the previous sections, while 
others, e.g., the differences owing to different evolutionary tracks, by
other authors (Fang et al. 2009).
Our estimate is that in general  \Macc\ is known with an uncertainty of a factor
3--5.

\begin{table}
\centering
\caption{Accretion properties of the class II objects and objects with transitional and 
evolved disks. }
\begin{tabular}{c c c c c c c}
\hline \hline
 Object & class & $\rm{M_{*}}$ & $\rm{L_{*}}$ & $\rm{T_{eff}}$ &  $\rm{logL_{acc}}$ & $\rm{logM_{acc}}$ \\
  & & ($\rm{M_{\odot}}$) & ($\rm{L_{\odot}}$) & (K) & ($\rm{L_{\odot}}$) & ($\rm{M_{\odot} \, yr^{-1}}$) \\
\hline
 SO299 &   TD &      0.35 &      0.14 & 3400 &     -2.93 &     -9.82 \\ 
  SO341 &    II &      0.60 &      0.39 & 3600 &     $<$ -2.26 &     $<$ -9.32 \\ 
  SO397 &    II &      0.30 &      0.18 & 3300 &     -2.73 &     -9.48 \\ 
  SO435 &    II &      0.20 &      0.09 & 3200 &     $<$ -3.38 &    $<$ -10.07 \\ 
  SO444 &  EV &      0.35 &      0.16 & 3400 &     $<$ -2.90 &    $<$ -9.78 \\ 
  SO462 &    II &      0.20 &      0.15 & 3200 &     -3.08 &     -9.67 \\ 
  SO482 &    II &      0.20 &      0.05 & 3200 &     -3.08 &     -9.91 \\ 
  SO485 &    II &      0.25 &      0.04 & 3300 &     -2.39 &     -9.37 \\ 
  SO490 &    II &      0.15 &      0.07 & 3100 &     -2.73 &     -9.32 \\
  SO500 &    II &      0.10 &      0.02 & 3000 &     -3.27 &     -9.92 \\  
  SO514 &    II &      0.13 &      0.04 & 3100 &     $<$ -3.88 &    $<$ -10.56 \\
  SO518 &    II &      1.00 &      0.41 & 4000 &     sat &     sat \\ 
  SO520 &    II &      0.30 &      0.15 & 3300 &    $<$ -3.02 &    $<$ -9.82 \\ 
  SO537 &    II &      0.07 &      0.02 & 3000 &     $<$ -4.32 &     $<$ -10.83 \\ 
  SO540 &    II &      1.00 &      0.43 & 4000 &     sat &     sat \\ 
  SO562 &    II &      0.40 &      0.22 & 3400 &     -1.80 &     -8.66 \\ 
  SO563 &    II &      0.40 &      0.51 & 3300 &     -2.30 &     -8.95 \\ 
  SO583 &    II &      0.75 &      1.52 & 3600 &     sat &     sat \\ 
  SO587 &  EV &      0.30 &      0.27 & 3300 &     $<$ -2.78 &     $<$ -9.45 \\ 
  SO598 &    II &      0.20 &      0.11 & 3200 &     -3.09 &     -9.75 \\ 
  SO615 &  EV &      0.60 &      0.69 & 3500 &     sat &     sat \\ 
  SO638 &    EV &      1.00 &      0.46 & 3900 &     sat &     sat \\ 
  SO646 &    II &      0.30 &      0.12 & 3300 &     -2.05 &     -8.89 \\ 
  SO657 &    II &      0.06 &      0.02 & 2900 &     $<$ -4.49 &    $<$ -10.91 \\ 
  SO662 &    II &      0.90 &      0.41 & 3900 &     $<$ -2.06 &     $<$ -9.26 \\ 
  SO663 &    II &      0.20 &      0.14 & 3200 &     $<$ -3.20 &    $<$ -9.82 \\ 
  SO674 &    II &      0.30 &      0.14 & 3300 &     $<$ -3.06 &    $<$ -9.88 \\ 
  SO697 &    II &      0.60 &      0.63 & 3500 &     -1.92 &     -8.76 \\ 
  SO700 &  EV &      0.08 &      0.01 & 3000 &     $<$ -4.46 &    $<$ -11.11 \\ 
  SO710 &    II &      0.50 &      0.30 & 3500 &     -2.28 &     -9.19 \\ 
  SO728 &  EV &      0.20 &      0.25 & 3100 &     $<$ -3.08 &     $<$ -9.53 \\ 
  SO736 &    II &      0.80 &      1.73 & 3700 &     sat &     sat \\ 
  SO738 &    II &      0.11 &      0.03 & 3100 &     $<$ -4.04 &    $<$ -10.7 \\ 
  SO739 &    II &      0.10 &      0.04 & 3000 &     $<$ -4.03 &    $<$ -10.54 \\ 
  SO750 &    II &      0.09 &      0.03 & 3000 &     -3.51 &    -10.03 \\ 
  SO759 &  EV &      0.30 &      0.19 & 3300 &     $<$ -2.97 &    $<$ -9.67 \\ 
  SO762 &    II &      0.13 &      0.05 & 3100 &     $<$ -3.82 &    $<$ -10.44 \\ 
  SO774 &    II &      1.00 &      0.47 & 4000 &     $<$ -1.93 &     $<$ -9.17 \\ 
  SO818 &   TD &      0.70 &      0.21 & 3700 &     -2.26 &     -9.45 \\ 
  SO827 &    II &      0.35 &      0.06 & 3400 &     -2.74 &     -9.82 \\ 
  SO844 &    II &      0.60 &      0.42 & 3600 &     -2.01 &     -8.95 \\ 
  SO848 &    II &      0.20 &      0.03 & 3300 &     -2.85 &     -9.86 \\ 
  SO859 &    II &      0.35 &      0.17 & 3400 &     -2.75 &     -9.61 \\ 
  SO865 &    II &      0.35 &      0.13 & 3400 &     -2.59 &     -9.50 \\ 
  SO866 &    II &      0.20 &      0.05 & 3200 &     -3.28 &    -10.12 \\ 
  SO897 &   TD &      0.80 &      0.50 & 3700 &     $>$ -2.07 &     $>$ -9.13 \\ 
  SO905 &  EV &      0.60 &      0.35 & 3600 &     $<$ -2.39 &     $<$ -9.38 \\ 
  SO908 &    EV &      0.20 &      0.10 & 3200 &     -2.68 &     -9.37 \\ 
  SO917 &  EV &      0.20 &      0.03 & 3300 &     $<$ -3.73 &    $<$ -10.71 \\ 
  SO927 &    II &      0.70 &      0.29 & 3700 &     $<$ -2.41 &     $<$ -9.53 \\ 
  SO936 &    II &      0.08 &      0.01 & 2900 &    $<$ -4.58 &    $<$ -11.19 \\ 
  SO967 &    II &      0.25 &      0.08 & 3300 &    $<$ -3.39 &   $<$ -10.25 \\ 
  SO981 &  EV &        1.3 &      1.17 & 4200 &     sat &    sat \\ 
  SO984 &    II &      0.90 &      0.53 & 3800 &     $<$ -2.12 &     $<$ -9.24 \\ 
 SO1009 &  EV &       0.15 &      0.02 & 3200 &      -1.97 &      -8.89 \\
 SO1036 &    II &      0.75 &      0.45 & 3700 &    $<$ -2.17 &    $<$ -9.22 \\ 
 SO1050 &    II &      0.20 &      0.08 & 3200 &     $<$ -3.49 &    $<$ -10.22 \\ 
 SO1057 &  EV &      0.15 &      0.08 & 3100 &    $<$ -3.67 &   $<$ -10.25 \\ 
 SO1059 &    II &      0.06 &      0.01 & 2900 &     $<$ -4.33 &    $<$ -10.86 \\ 
 SO1075 &    II &      0.17 &      0.13 & 3100 &      $>$ -1.66 &    $>$ -8.18 \\
 SO1156 &    II &      1.00 &      0.63 & 3900 &     sat &     sat \\
 SO1182 &    II &      0.20 &      0.10 & 3200 &     -3.12 &     -9.79 \\ 
 SO1193 &    II &      0.11 &      0.03 & 3100 &     -3.32 &    -10.01 \\ 
 SO1230 &    II &      0.20 &      0.08 & 3200 &     -3.32 &    -10.06 \\ 
 SO1260 &    II &      0.35 &      0.13 & 3400 &     -1.68 &     -8.60 \\ 
 SO1266 &    II &      0.16 &      0.07 & 3100 &     -3.47 &    -10.09 \\ 
 SO1268 &   TD &      0.09 &      0.02 & 2900 &     $<$ -4.49 &    $<$ -11.06 \\ 
 SO1274 &    II &      0.70 &      0.64 & 3600 &     $>$ -2.15 &  $>$ -9.07 \\ 
 SO1323 &  EV &      0.20 &      0.07 & 3200 &     $<$ -3.51 &    $<$ -10.25 \\ 
 SO1327 &    II &      0.30 &      0.20 & 3300 &     -2.66 &     -9.39 \\ 
 SO1361 &    II &      0.55 &      0.53 & 3500 &     $>$ -1.82 &     $>$ -8.65 \\ 
 SO1362 &    II &      0.15 &      0.09 & 3100 &     -3.07 &     -9.61 \\ 
 
\hline
\end{tabular}
\label{paramCII}
\end{table}

\section {Results}
\label{result}

Detectable values of accretion luminosity and mass accretion rate were obtained 
for 42\% of disk objects. Another 42\% have upper limits in  
accretion luminosity and mass accretion rate. The remaining 16\% 
have lower limits (see \S 4.3).

Gatti et al.~(2008)  compute \Lacc\  
from the luminosity of the hydrogen recombination line Pa$\gamma$, 
 following the procedure described by Natta et al.~(2006). 
 
\begin{figure}[t]
\includegraphics[width=9cm]{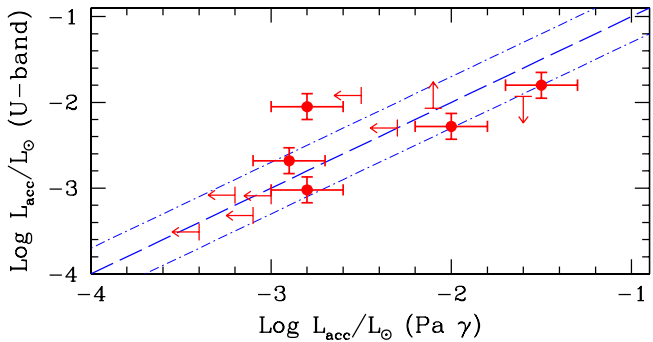}%
\caption{ Comparison between \Lacc\ computed from Pa$\gamma$ (Gatti et al.~2008) and \Lacc\ 
obtained from U-band photometry for  stars in common. The dashed line refers to the 
same value of \Lacc\ , the dashed-dotted lines to $\pm 0.5$ in $\log$ \Lacc.
}
\label {GattiCOM}
\end{figure}

Figure~\ref{GattiCOM} compares the values of \Lacc\ derived with the two different methods 
(U-band excess emission and Pa$\gamma$ recombination line respectively) 
for the 12 stars we have in common; it agrees well. 
Another nine stars of the Gatti et al.~(2008) sample are class II objects and were 
observed with {\it{Spitzer}}. 
We could not observe these stars either because they are outside the FORS fields, 
or because they are close to the 
brighter stars of the \sori\ quintuplet system. 
We added these nine stars to our sample.
In Table.~\ref{GATTItab} the properties of the sub-sample derived from 
Gatti et al.~(2008) are listed. 
For these stars we recomputed the stellar properties adopting the Baraffe et al.~(1998) 
evolutionary tracks., in the same way as for the present sample. \\

\begin{table}
\centering
\caption{Accretion properties of the class II objects taken from Gatti et al.~(2008). 
The masses and luminosities of the listed stars have been recomputed 
according the Baraffe et al.~(1998) evolutionary tracks.}
\begin{tabular}{c c c c c c c}
\hline \hline
 Object & class & $\rm{M_{*}}$ & $\rm{L_{*}}$ & $\rm{T_{eff}}$ &  $\rm{logL_{acc}}$ & $\rm{logM_{acc}}$ \\
  & & ($\rm{M_{\odot}}$) & ($\rm{L_{\odot}}$) & (K) & ($\rm{L_{\odot}}$) & ($\rm{M_{\odot} \, yr^{-1}}$) \\
\hline
  SO451 &  II &      0.30 &      0.12 & 3350 &  -2.20 &  -9.10 \\ 
  SO682 &  II &      0.60 &      0.30 & 3575 &     $<$ -2.50 &    $<$ -9.30 \\
  SO694 &  II &      0.15 &      0.10 & 3075 &     $<$ -3.20 &    $<$ -9.80 \\  
  SO723 &  II &      0.25 &      0.10 & 3250 &  -2.00 &  -8.90 \\
  SO726 &  II &      0.60 &      0.42 & 3575 &  -1.80 &  -8.60 \\  
  SO733 &  II &      0.45 &      0.36 & 3425 &  -2.00 &  -8.70 \\  
  SO871 &  II &      0.45 &      0.10 & 3500 &     $<$ -3.20 &    $<$ -10.30 \\ 
 SO1152 &  II &      0.60 &      0.30 & 3575 &     $<$ -2.40 &    $<$ -9.30 \\  
 SO1248 &  II &      0.13 &      0.13 & 3000 &     $<$ -2.80 &    $<$ -9.30 \\ 
 
\hline
\end{tabular}
\label{GATTItab}
\end{table}

\begin{figure}[t]
\includegraphics[width=9cm]{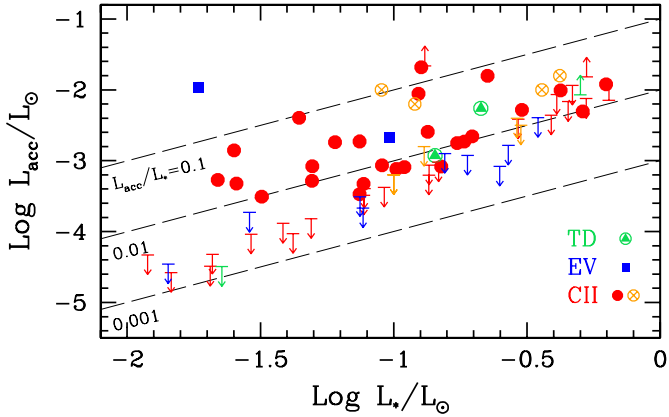}%
\caption{Accretion luminosity as a function of the stellar luminosity 
for class II, EV, and TD disks, as labeled.
The dashed lines show  $L_{acc}/L_{star}=$ 0.001, 0.01, and 0.1, respectively. 
The crosses surronded by a circle are the class II stars included here 
from the Gatti et al.~(2008) sample, and listed in Table~\ref{GATTItab}. 
Different colors for lower and upper limits can be distinguished 
in the online version of the  journal.
}
\label {LaccLstar}
\end{figure}

\begin{figure}[t]
\includegraphics[width=9cm]{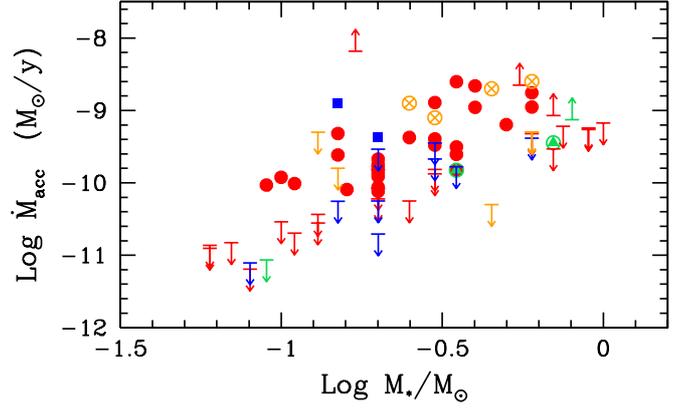}%
\caption{Mass accretion rate as function of stellar mass.
Symbols as in Fig.~\ref{LaccLstar}. 
(A color version of this figure is available in the online journal.)}
\label {MaccMstar}
\end{figure}

\begin{figure}[t]
\includegraphics[width=9cm]{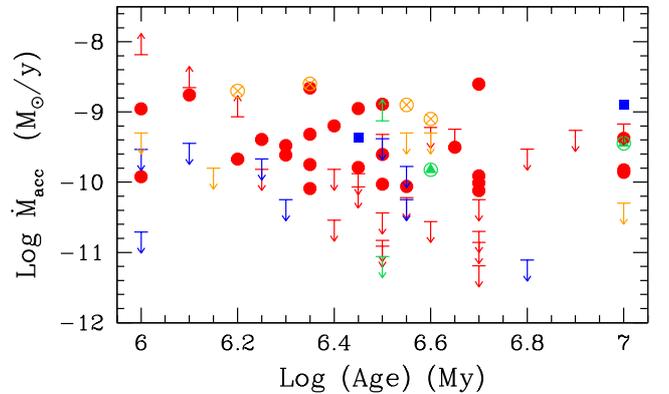}%
\caption{Mass accretion rate as function of the age of the stars.
Symbols as in Fig.~\ref{LaccLstar}. }
\label {MaccAGE}
\end{figure}

\begin{figure}[htbp]
\includegraphics[width=9cm]{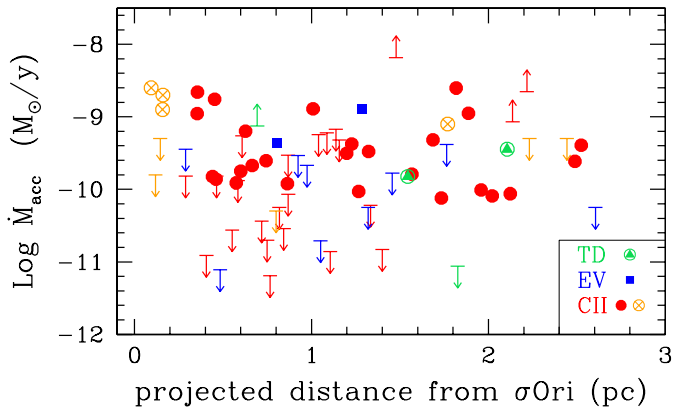}%
\caption{Mass accretion rates versus 
  projected
distance from the bright stars \sori\ in parsec. 
(A color version of this figure is available in the online journal.)
  }
\label {dist}
\end{figure}

Figures ~\ref{LaccLstar} - \ref{dist} show the relations between 
the accretion properties and different physical and 
morphological properties of the observed sample. \\

In Fig.~\ref{LaccLstar} we specifically plot the accretion luminosity 
as a function of the stellar luminosity 
for all class II objects; transitional and evolved disks are shown by different
symbols. 
The figure shows that very few stars have $L_{acc}$ larger than 0.1 $L_{star}$, and that 
most of them have values well below this limit. 
The accretion luminosities for the detected sources range mainly from 
$\sim10^{-2}L_{\odot}$ to $\sim10^{-4}L_{\odot}$. 
For any given \Lstar\ there is a large range of measured \Lacc\ that does not 
seem to vary significantly with \Lstar\ . \\

Figure~\ref{MaccMstar}  shows the mass accretion rate of class II stars 
as function of $M_{*}$. 
The data show a clear trend of increasing  $\dot M_{acc}$ with increasing $M_{*}$. 
Including upper and lower limits as actual detections, we find 
${\dot M_{acc}} \propto M_{*}^{1.6\pm0.4}$ with ASURV 
(Astronomy Survival Analysis Package, Feigelson \& Nelson~1985). 
The trend is confirmed using two different methods (the EM algorithm and the BJ algorithm) 
within the ASURV package. 
The slope became flatter when we excluded the upper and lower limits in the analysis, 
but remained still within the uncertainties.
From this plot we can clearly also see the large spread in ${\dot M_{acc}}$ 
(about two orders of magnitude)
for 
any value of $M_{*}$. 

In Fig~\ref{MaccAGE} we plot the mass accretion rates versus the age of the stars. 

Finally,
in Fig~\ref{dist} we plot the mass accretion rates as function of the projected distance 
from the central and bright O9.5 star \sori. 
We do not find any correlation between the mass accretion rates or
the disk morphologies and the 
distance from  \sori; it seems that the vicinity of the O9.5
star does not affect the disk properties significantly (see also Hernandez et al. 2007).
Rigliaco et al. (2009) studied
in detail one object
(SO587), which has a projected distance of 0.3 pc; based
on literature U-band photometry and optical spectroscopy, they
proposed that its disk is in the process of being photoevaporated, either
by \sori\ or by its own central star.
There are six additional  objects with a 
projected distance of$\simless$0.3 pc from \sori\ with a \Macc\ range of more
than a factor of ten, and it would be interesting to obtain high-resolution
optical spectra to study  wind diagnostics such as the optical forbidden lines
of [SII], [NII] and [OI].

We caveat that in both Fig.~\ref{MaccAGE} and Fig.~\ref{dist} we did not divide stars in mass bins
(which would result in too low number statistics); 
thus the relationship of \Macc\ vs. \Mstar\ seen in Fig.~\ref{MaccMstar} might mask possible trends 
between \Macc\ and age or \Macc\ and the projected distance from $\sigma$~Ori.

\section {Discussion}
\label{discussion}

\subsection {\Macc\ as function of \Mstar\ and time}

The distribution of mass accretion rates in a star-forming region
traces the physical processes that control disk formation and evolution
over the lifetime of the region, as well as the initial conditions, i.e.,
the mass and angular momentum distribution of the
molecular cores from which the stars form. It is 
a snapshot in time, which needs to be compared to models that follow
disk formation and evolution up to the age of the 
region (e.g., Dullemond et al.~2006;
Vorobyov \& Basu 2008, 2009).

\begin{figure}[t]
\centering
\includegraphics[width=9.0cm]{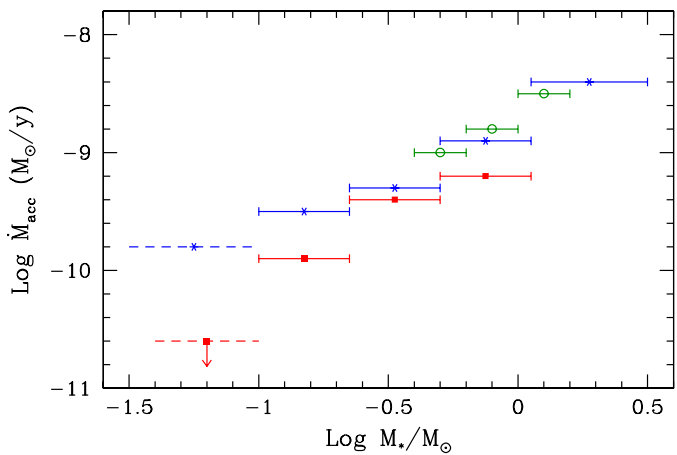}
\caption{
Median values of \Macc\ as function of \Mstar. Red lines (with a central square) refer to
\sori\ Class II and TD disks; the lowest mass bin should be considered as
an upper limit only (see text). Blue lines (crosses) show the distribution for \roph\ 
(data from Natta et al.~2006, see Appendix A);
green lines (circles) for Tr 37 (data from Sicilia-Aguilar et al.~2010). 
(A color version of this figure is available in the online journal.)}
\label {medians}
\end {figure}

It has been known for some years that
mass accretion rates  increase on average with
the mass of the central object.  In \sori, 
a logarithmic linear correlation over the whole mass range, 
\Macc$\propto M_{*}^{1.6\pm 0.4}$ (Sec.~\ref{result}), 
is similar within the uncertainties to what has found in Taurus (Calvet et al.~2004) 
and Ophiuchus (Natta et al.~2006), but flatter
than in L1630N and L1641 (Fang et al.~2009).  
However, in \sori\ 
the trend seems to be flatter for higher mass stars than for lower mass
objects, as also noted by Vorobyov \& Basu~(2008) in a compilation 
of all known \Macc\ values in various
star-forming regions.
This is  shown  in
Fig.~\ref{medians}, which plots
median values of \Macc\ as function of  \Mstar\ for
Class II and TD disks.
All \Mstar\ intervals above 0.1 \Msun\ contain roughly the same number of stars
(12--20). 
The values plotted in Fig.~\ref{medians} treat upper and lower
limits as actual detections; if we make the alternative assumption that lower
(upper) limits
are all smaller (larger) than the lowest (highest) measured
value in the bin, the changes are very small;
however, the large number of  limits and the
small number of objects in each bin make it meaningless to derive values for the
upper and lower quartiles, to characterize the spread of \Macc\ seen in Fig.\ref{MaccMstar}.
The mass interval below 0.1 \Msun\ contains only two detections and  six upper limits to \Macc, 
and the value in Fig.~\ref{medians} (Log \Macc=-10.8 \Msun/y) should be considered as an
upper limit to the median. 
The deficit of relatively strong accretors among
the \sori\ BDs is confirmed by  the analysis of
a much larger (40 objects, see Appendix B), 
optically selected sample of very low-mass stars and BDs 
(Lodieu et al.~2009), included in our U-band survey, 
which contains only five objects with  Log \Macc $>-10.8$ \Msun/y, 
two of which are 
also in the Hernandez et al.~(2007) {\it Spitzer} sample.

The flattening of the \Macc\ vs. \Mstar\ relation at higher
masses is very clear.
Vorobyov \& Basu (2009) compute numerical models of the collapse of
a distribution of
prestellar cores that include the formation and evolution of
circumstellar disks and follow it for 3 My, roughly  the age of \sori.
Their models
predict a flattening of the \Macc\ vs. \Mstar\ relation for
stellar masses higher than about 0.3 \Msun, due to the effect
of gravitationally induced torques in the early stages
of the evolution after the formation of the central star. These gravitational
instabilities have little effect on lower-mass objects, where viscous
evolution dominates at all times.
Although the \Macc\ values predicted by  Vorobyov \& Basu~(2009) are
somewhat higher than the observations, the \sori\ results  definitely support 
their models 
and the importance of self-gravity in the early evolution
of more massive disks, already suggested by, e.g., Hartmann et al. (2006). 

The Vorobyov \& Basu (2009) models include only viscous evolution and
gravitational instability; other physical processes may occur during
\sori\ lifetime, such as
photoevaporation and planet formation, leading to  disk
dissipation on shorter time scales.
This may be a selective process, if, as 
indicated by the statistics of IR-excess emission for stars
of different mass in different star-forming
regions, disk dissipation occurs faster in more massive stars.
In \sori, for example, Hernandez et al.~(2007) estimate a fraction of objects with disks
that increases from $\sim 10$\% for Herbig Ae/Be stars to $\sim 35$\% for
T Tauri stars and BD candidates.
If so, the comparison of the observations with the model predictions needs
to be taken with care.

A way to investigate the relative importance of these different processes
is to compare the statistical properties of  the distribution of
\Macc\ on \Mstar\ for regions of different age.
This approach is limited at the moment because to the
best of our knowledge, there are only two
other suitable samples:
\roph\ (Natta et al.~2006) and Tr 37 (Sicilia-Aguilar et al.~2010). 
In other cases,
no mass accretion rates are available
for complete sample of Class II objects (as in Taurus), 
or upper limits to \Macc\ are not
provided (as in the two Orion regions studied by Fang et al. 2009), which makes
statistical studies very difficult.

We have computed  median values of \Macc\ in \roph\ from
the results of Natta et al. (2006), which we revised to take into account the new
estimates of the distance  that were recently 
published (see Appendix A for details).
The  \roph\ sample covers a mass range between $\sim$ 0.03 and 3 \Msun;
the median values of \Macc\ are shown in Fig.~\ref{medians};
the figure also shows the
results for the older (about 4 My), more distant region Trumpler 37, for which
Sicilia-Aguilar et al.~(2010) provide values of \Macc\ derived from U-band
photometry for a {\it Spitzer}-selected sample of stars.
The stars cover the mass range 0.4-1.6 \Msun\ ,
based on the Siess et al.~(2000) evolutionary tracks. 

For \Mstar\ roughly larger than 0.2 \Msun, the three regions 
have similar values of the median \Macc\ within the
uncertainties in spite of their
difference in age (more than a factor of 3). 
If the disk properties at a very early stage were
the same, this would imply a slower time evolution of \Macc\ than predicted by
disk  models (see, e.g., Hartmann et al. 1998; Dullemond et al. 2006; Vorobyov \& Basu 2009) and confirms the result of
Sicilia-Aguilar et al.~(2010), based on a small sample of stars for which
individual ages could be estimated.
The difference between \Macc\ medians increases as \Mstar\ decreases, 
and becomes very large (factor of 10 at least) for \Mstar\ $<0.1$ \Mstar\ between \roph\ and \sori. 
This is consistent with the predictions of viscous models if the original disk 
properties in the two regions are the same and no disk around these
very low-mass stars dissipates within, e.g., 3 My.

If indeed
the apparent slow time evolution of solar mass stars
is a result of the continuous
loss of the less massive,  
for the lower accreting disks 
the two different effects (namely
the decrease of \Macc\ with time, and dissipation of disks 
once they fall below a critical \Macc\ value) roughly compensate, so that
median \Macc\ changes little.
If that is true, one should find that the lowest values of \Macc\ do not vary with time, 
because they are fixed by disk dissipation, while the highest  values
decrease as expected from viscous evolution. There is a hint that this is indeed the case 
(seel also L1641 and L1630; Fang et al. 2009), but the statistics is poor
and the definition of the upper envelope to the \Macc\ values too
uncertain for the moment.

\subsection {Mass accretion rates vs disks morphologies}
\label{SEDmorph}

\begin{figure}[!h]
\centering
\includegraphics[width=8.5cm]{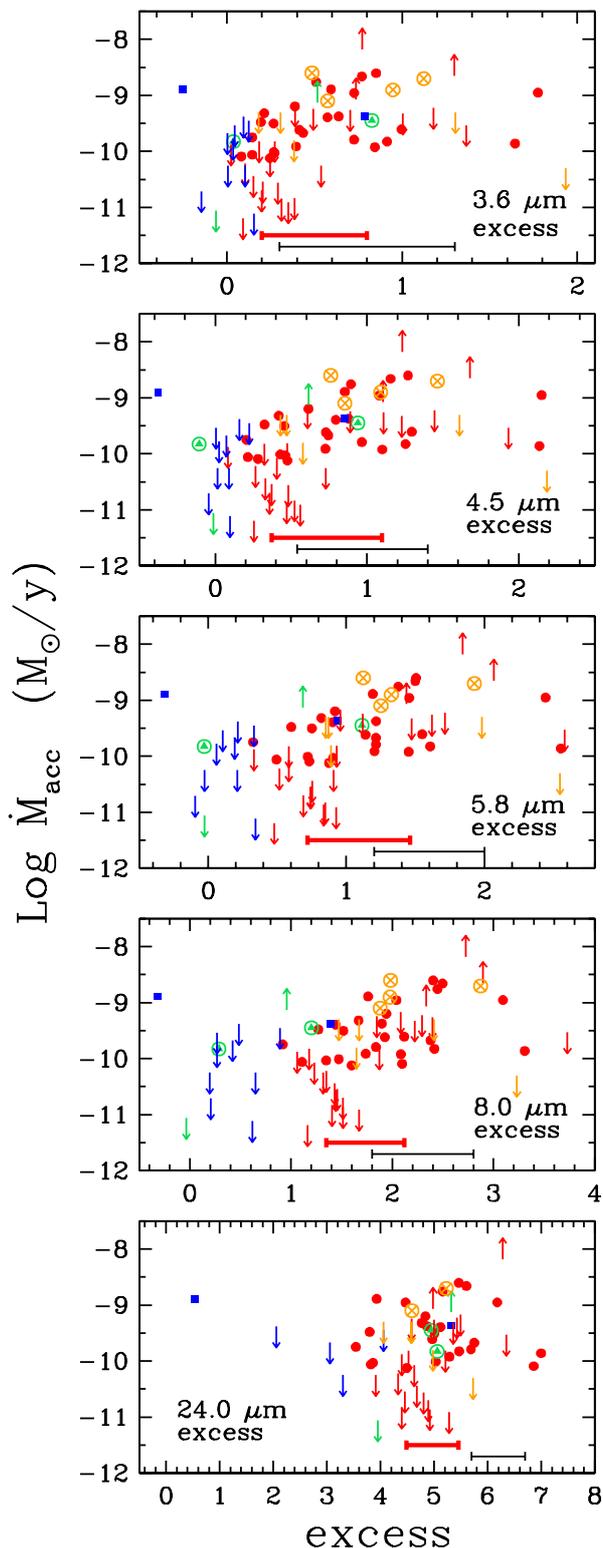} 
\caption{ \Macc\ vs excess emission (in mag) in the four IRAC {\it Spitzer} bands 
and for the 24\um\ MIPS band. 
The median excess emission and first quartiles are shown in each panel
by the red (thick) horizontal lines; for comparison, we also plot
 the  median Taurus class II SED
(black (thin) line; D'Alessio et al. 2006).
Symbols as in Fig.~\ref{LaccLstar}. 
(A color version of this figure is available in the online journal.)
}
\label {excess}
\end{figure}

Fig.~\ref{excess} plots the mass accretion rates for the \sori\ sample as function of
the excess over the photospheric emission for the four IRAC bands
and for the 24 $\mu$m MIPS band. 
The excess emission is defined as the difference between the
observed  ({\it Spitzer}-I) color  and that of Class III stars of the same effective temperature, assuming that no excess is present in the I band.
In each panel, the horizontal bars show the  median of the
excess distribution for Class II disks
and the lower and upper first quartile; for
comparison, we also show the lower and
upper first quartile of the excess for the Taurus median SED of classical disks
(D'Alessio et al.~2006).

The median of the excess emission distributions is lower in \sori\ than in
Taurus, as expected if on average \sori\ stars are older  than Taurus objects,
suggesting a higher degree of grain settling in older regions (e.g., Hernandez et al.~2007).
Note  that the \sori\ sample includes lower-mass objects than the Taurus one,
and that disk models predict lower excess fluxes for BDs in this range of
wavelengths; however, we do not find any correlation of the measured excess
with the mass of the star, and we tend to exclude that the difference between
the two regions is  due to the different mass range only.
The figure also shows the location of the 11 EV disks  with
a determination of \Macc\ and the 4 TD; for six of them we also
show complete SEDs and \Ha\ profiles in Appendix C.

There is no statistically
significant  correlation of \Macc\ with the excess emission. However,
there are a few aspects of these plots that can be understood
if, as expected in viscous models, \Macc\ traces the
surface density of the inner disk.
Objects with 
\Macc $\sim 10^{-10}- 10^{-9}$ \Msun/y are distributed
over the whole range of excess values.
This is expected in optically thick disks, where the observed emission
depends on inclination, inner radius and degree of flaring, but not
on the actual disk surface density.
Objects with  \Macc$\simless 10^{-10}$ \Msun/y 
 tend to have very low excess in  all
{\it Spitzer} bands. We think that most of them have
in fact optically thin disks, which are characterized
by much lower emission, roughly proportional
to the surface density, i.e., to \Macc.
The value \Macc $\sim 10^{-10}$ \Msun/y is a reasonable
threshold
for the transition from optically thick
to optically thin (inner) disks
(D'Alessio et al. 2006).

A last point to note is that disks with \Macc $\simgreat 10^{-9}$ \Msun/y
have all large excess emission (in the upper
quartile of the distribution), and indeed the upper
envelope of the \Macc\ distribution seems to correlate with
the amount of excess emission. Given the small number
of objects in the high range of \Macc, the significance of
this is unclear. If true, it would suggest a very interesting
relation between \Macc\ and grain growth and settling, i.e.,
processes that can change the grain opacity and the disk
flaring.

The potential of plots like Fig.~\ref{excess} in constraining disk model parameters
should  be exploited further. However, this is well beyond the scope of this
paper.

Fig.~\ref{excess} also shows the location of the 11 EV disks with an \Macc\ determination.
Nine of them have upper limits to \Macc, some well below $10^{-10}$ \Msun/y, and very small excess emission, generally below the lower quartile of the Class II
distribution. From the upper limits
to \Macc, it is likely that at least half of them
are indeed optically thin disks.
There are two exceptions, one (SO908) is probably a missclassified
Class II object,
as it shows significant excess emission in all bands and broad \Ha\ emission,
consistent with its measured accretion rate (\Macc$=4\times 10^{-10}$ \Msun/y)
(see  Appendix C).
The other EV object (SO1009) has one of the highest accretion rates in our sample
($1.2\times 10^{-9}$ \Msun/y) and no detectable excess emission in all bands;
indeed, its classification as a disk object is dubious. 
No additional data are available in the literature, and we may
have detected a strong chromospheric flare. It would be interesting
to monitor this object further.

The location of the four TD on Fig.~\ref{excess} is also shown. They
are scattered through the plot:
two (SO818
and SO897) are consistent (both in \Macc\ and  excess emission, see
also Appendix C)
with optically thick inner disks. One (SO1268) has very
likely a very optically thin
inner disk, with a very low upper limit
to \Macc ($\sim 9 \times 10^{-12}$ \Msun/y)
and small excess emission even at 24 $\mu$m.
The fourth TD object  (SO299) has an accretion
rate typical for its mass, 24 $\mu$m excess as Class II of similar
accretion rate, but negligible excess
in the IRAC bands.
The large spread of properties of TD confirms the analysis of 
Muzerolle et al. (2010) and Sicilia-Aguilar et al. (2010) and
their conclusions that very likely the "transitional" SEDs trace a variety of
different physical situations.

\section {Conclusions}

We reported the results of a U-band survey with FORS1/VLT
of a large area in  
the \sori\ star-forming region. We combined the U-band photometry with
literature results to compute accretion luminosity and mass accretion rates
from the U-band excess emission for all objects detected by
{\it Spitzer} in the FORS1 field  and classified by
Hernandez et al. (2007)  as likely members of the cluster.
In total, there are 72 objects with evidence of a disk
from near- and mid-IR photometry and 115 class III (diskless) stars.
Among the disk objects, four (out of the seven identified by Hernandez et al.~2007)
are transitional disks, and 14 are evolved disks.
We derived the photospheric parameters of all stars from the existing
V, R, I, and J photometry and used the U-$\lambda$ colors of class III as
templates for the photospheric and possible chromospheric emission.
Our final sample, for which we provide estimates of the mass accretion rates,
contains 58 Class II 
(49 class II stars for which we derive the accretion properties 
from the U-band excess emission, 
and nine stars with accretion properties from literature, for which we checked 
the consistency with our results), 
four TD and 11 EV disks, over a mass range of 
between  $\sim 0.06 $ and  $\sim 1.2 M_{\odot}$.

We analyzed the behavior of \Macc\ as function of mass and age of
the individual stars, of the properties of the IR SED, and
of the distance from the bright star \sori.

There is no correlation of \Macc\ with the distance from \sori,
confirming that the effect of the O9.5 star on its surroundings is not
strong (see also Hernandez et al.~2007); however, our sample does not include
objects with projected distances smaller than 0.1 pc.

We find a strong relation between \Macc\ and \Mstar, with a very large
spread of \Macc\ values for any given \Mstar, similar to other
star-forming regions 
(Calvet et al.~2004; Natta et al.~2006; Fang et al.~2009; Sicilia-Aguilar et al.~2010). 
If fitted with a linear correlation, the slope is $1.6 \pm 0.4$.
As noted by Vorobyov \& Basu~(2009) for a compilation of
accretion rates in different star-forming regions, a linear fit
is not the best description of the data. We computed 
median \Macc\ values and showed that the relation between \Macc\ and \Mstar\
is flatter at higher masses and steepens significantly for very low-mass
stars and BDs.  Such a trend is predicted by models of core collapse and disk
evolution that include viscosity  and gravitational instabilities
(Vorobyov \& Basu~2009), which control the evolution of  more massive disks.

These models follow the disk evolution to the age of \sori, under the assumption
that disks evolve only by  accreting onto the star. However,
other processes such as photoevaporation 
and/or planet formation, may cause disk dissipation 
(e.g. Hollenbach~2000; Dullemond et al.~2007). This may be
a selective effect, which affects  higher mass stars more than lower mass stars,
changing the observed dependence of the \Macc\ distribution with \Mstar.
We have compared the \Macc--\Mstar\ distribution in \sori\
to that of the two other star-forming regions in the literature for which
an IR-selected sample is available, namely \roph\ (Natta et al.~2006 and
Appendix A) and Tr 37 (Sicilia-Aguilar et al.~2010).
The comparison indicates that the median \Macc\ values for
higher \Mstar\ are closer 
than predicted by simple viscous models, suggesting that
selective disk dissipation may be important. However, we note
that the significance of this comparison is not very strong, because 
the results for \roph\ are uncertain and the number of objects
in each mass bin is not large. Moreover, the number and distribution
of upper limits to \Macc\ affects
the determination of median values in some
mass ranges and prevents us from deriving
upper and lower quartiles of the distributions. 

The behavior of \Macc\ as function of the excess emission
in the {\it Spitzer}  bands suggests that at \Macc $\sim 10^{-10}$ \Msun/y
the (inner) disks change from  optically thin  to optically thick.
Objects with \Macc\ in the range $\sim 10^{-10} - 10^{-9}$ \Msun/y span the whole
range of observed excesses, from very low to very large. 
Objects with \Macc $\simgreat 10^{-9}$ \Msun/y (the largest values
in the sample) all have large IR excess. 
Viscous disk models (e.g., D'Alessio et al. 2006) predict that for 
\Macc =$10^{-10}$ the surface density  at 1 AU from a T Tauri star will be on the order of 
 1-10 g cm$^{-2}$, depending on the grain properties and dust settling.   
The emission of these disks in the {\it Spitzer} bands will be optically thin, 
unless grains are sub-micron size.
The trend we tentatively observe among optically thin and optically thin disks, 
which needs to be confirmed in larger samples, may indicate
a link between the mass accretion rate and the
grain properties, which in turn control the disk geometry,a connection that is worth 
to be further explored.

The four TD stars included in our sample seem to cover a variety of properties,
and only one of them has a mass accretion rate as Class II of similar mass,
and negligible excess
emission to wavelengths $> 8 \mu$m. The other three could not be distinguished from Class II objects in the \Macc\ vs. excess emission plots.
We can only agree with the conclusions that the "transitional" properties of the
SEDs are likely caused by a variety of different properties (Muzerolle et al.~2010; Sicilia-Aguilar et al.~2010).

\begin{acknowledgements}
We thank Fabrizio Massi for useful suggestions on the data reduction and Nicolas Lodieu 
for providing the data on the \sori\ brown dwarfs. 
This publication is based on observations made with FORS1@VLT and SARG@TNG. 
We acknowledge the staff of the ESO Data Management and Operations department, 
who performed our observations in service mode.
\end{acknowledgements}

\appendix

\section{$\rho$ Ophiucus}
\label {AppendixOph}

\begin{figure}
   \centering
   \resizebox{\hsize}{!}{\includegraphics[]{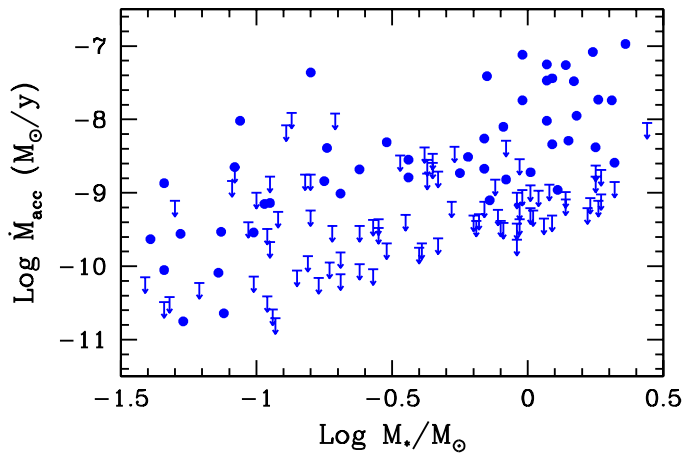}}
        \caption{Mass accretion rate versus stellar masses for the star of the \roph\ sample. 
         The stellar parameters and the accretion properties were recalculated 
         assuming a distance of 130 pc 
         (Lombardi et al.,2008) 
         instead of 160 pc assumed by Natta et al.~(2006), and 1 My age.
        }
\label {MaccMstarOPH}
\end{figure}

The  \roph\ sample is particularly interesting, because it
is similar to the \sori\ one in being  an IR-selected sample of class II,
complete to a limiting mass of about 0.05 \Msun\ (Bontemps et al.~2001).
Natta et al.~(2006) computed mass accretion rates from the luminosity of
Pa$\beta$ assuming
a distance of 160 pc;  since no extensive spectral type determinations were
available, they followed the method outlined by Bontemps et al.~(2001)
and derived the stellar parameters assuming coeval star formation 
at 0.5 My and the evolutionary tracks of D'Antona \& Mazzitelli~(1998; the isochrone method).
They found that the mass accretion rates could be fitted by a linear relation
$\propto M{*}^{1.8\pm 0.2}$ over a mass interval 0.03--3 \Msun, 
with a very large 
spread for any given \Mstar. 
However, the Natta et al.~(2006) results need to be
reconsidered, since new measurements of the \roph\ distance  
yield considerably low values,
120-130 pc (Lombardi et al. 2008; Loinard et al. 2008; Snow et al. 2008).
The corresponding decrease in luminosity implies an older age
for the region.
We have redetermined stellar parameters and 
mass accretion rates for all objects in Natta et al. (2006), adopting a 
distance of 130 pc and
evolutionary tracks of Baraffe et al. (1998) for 1 My.
The new values of
\Macc\ are somewhat lower than previous ones,
and \Mstar\ higher, especially for higher masses (see Fig.~\ref{MaccMstarOPH}). 

As a consequence, if fitted with a single power-law,
the correlation between \Macc\ and \Mstar\ is flatter, with a slope of 
$1.3 \pm 0.2$.
The new distance, the older age and the different evolutionary tracks
contribute to this result. In particular, the 
dependence of the \Macc\ -- \Mstar\ relation on
the adopted evolutionary tracks is well known (see Fang et al. 2009)
and is particularly strong in \roph, given the method
used to determine the stellar parameters.

\section{BD and very low-mass stars from the Lodieu et al. photometric survey}
\label {AppendixA}

An independent sample of very low-mass stars and brown dwarfs
has been selected from the list of \sori\ members and candidate members of Lodieu et al. (2009).
We applied a first selection criterium on the z,(z-J) diagram
computed \Teff\ by comparing observed
z, Y, J colors to synthetic ones from 
the theoretical models of Baraffe et al. (1998) for $\log g$=4.0
and luminosities from the observed J mag and model-predicted bolometric corrections.
We then performed a further selection based on the location of the 
objects on the HR diagram, excluding all stars 
with \Mstar$>$0.13 \Msun. Our sample of candidate young very low-mass stars and BDs in \sori\ 
is then formed by 80
objects,  40 of which were
included in the U-band FORS1 survey.

{\it Spitzer} IRAC detections exist for 21/40 objects;
three of them are uncertain members according to
Hernandez et al.~(2007).  Of the 21, six are classified as class II,
one is a transitional disk, one an evolved disk and 13 are class III objects.
The 18 confirmed members  are included in the sample analyzed in the main text of the paper. 
Note that the determination of the stellar parameters, \Teff\ in 
particular is performed using different photometric bands with respect to the bands used in 
the main text;
the difference in \Teff\ are in general
within the uncertainties discussed in \S 3.1;
however,
some of the  objects in Table 2, although included in the
Lodieu sample, were not 
selected with the criteria applied here.

\begin{figure}
   \centering
   \resizebox{\hsize}{!}{\includegraphics[]{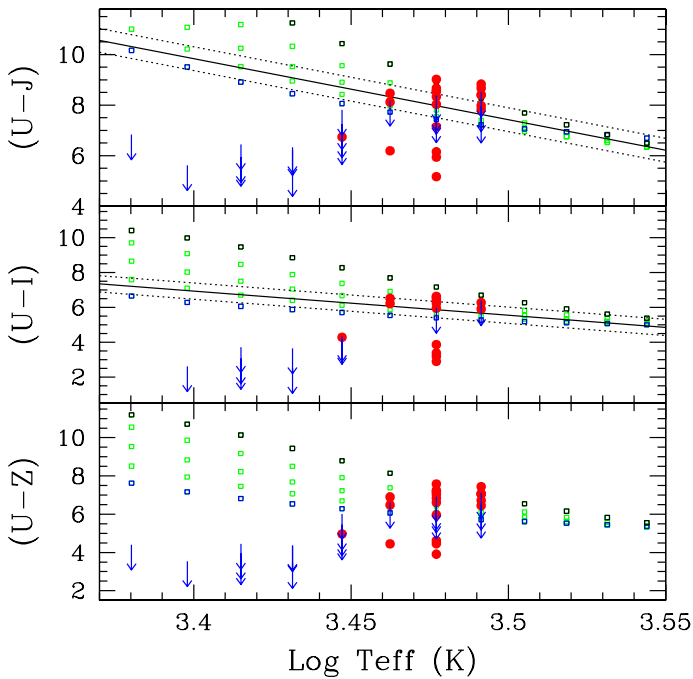}}
        \caption{ U-$\lambda$ colors as function of \Teff. The red circles are 
objects with measured U-band fluxes, arrows are objects with U-band upper
limits. The squares are model-predicted colors for different gravity, from
5.5 (black squares, top) to 3.5 (blue squares, lowest). 
The solid lines show the best-fit relations for Class III derived in Sec.4.1,
extrapolated to lower \Teff; dashed lines are $\pm 2\sigma$. We will consider 
as accretors objects with colors below the photospheric strip: five BDs have clear evidence of U-band excess 
(note that for two we only have two colors), one only
marginal evidence (from (U-I) and (U-J), while (U-Z) is photospheric). 
Hereafter we mark the upper limit as dashes for clarity.
(A color version of this figure is available in the online journal.)}
\label {ucolors}
\end{figure}

Of the 40 objects in our sample, 20 have been detected in the U-band, while 
for the other 20 we have upper limits only.
We derived the accretion luminosity and mass accretion rates
as in \S 4.3.
The calibration of the photospheric colors (U-J) and (U-I)
as function of \Teff\
using our U-band photometry of Class III objects extends to \Teff$\sim 2900$ K (Sec.4.1);
we compare it to
 synthetic colors from the
Baraffe et al. (1998) model atmosphere 
to extend the relations to lower \Teff. The Class III colors agree well with the models; 
the results are shown in Fig.~\ref{ucolors}.

\begin{figure}
   \centering
   \resizebox{\hsize}{!}{\includegraphics[]{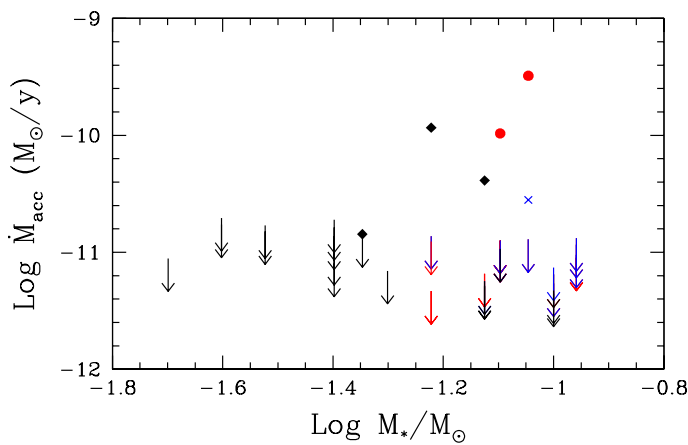}}
        \caption{ Mass accretion rate as function of \Mstar. Dots are class II objects
with measured \Macc, the cross is a class III stars with measured \Macc, and diamonds refer to object with no 
{\it Spitzer} data. Arrows are objects with upper limits (both U-band detections and non-detections). 
Colors (only in the on-line version) indicate the
SED Class : red for class II, blue for class III; black for objects with no {\it Spitzer}
data. 
(A color version of this figure is available in the online journal.)}
\label {macc}
\end{figure}

Fig.~\ref{macc} shows \Macc\ vs. \Mstar;
we find that only five objects (all with \Mstar$\>0.06$ \Msun) have \Macc$>-10.8$,
the median in \S 6.1. 
of these, two are class II sources (SO500 and SO848, also in Table 2), 
one is a class III
(SO641, possibly misclassified), three have no Spitzer
detections.
No other object with higher values of \Macc\ is detected.

\section{SEDs and H$_\alpha$ of EV and TD objects}
\label {AppendixB}

\begin{figure}[htbp]
\includegraphics[width=9cm]{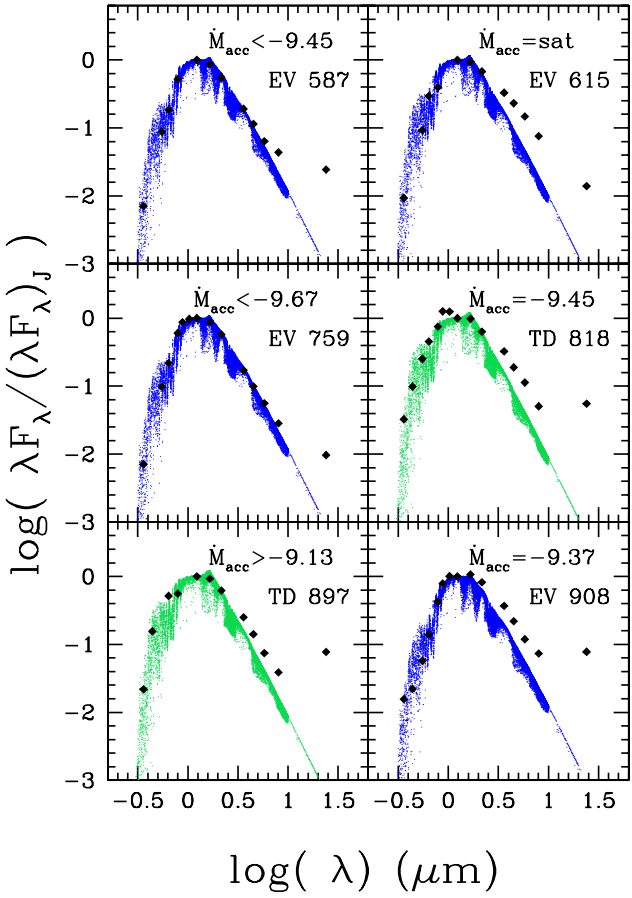}%
\caption{SEDs of four EV disks and 2 TD. The squares shows the 
observed fluxes (see Table~\ref{phot}). The lines plot the model atmosphere from
Allard et al. (2000) at the appropriate \Teff, $\log g=4.0$, normalized to the J band. 
Each panel gives the name of the star (see Table~\ref{AccPROP}) and the mass accretion rate.}
\label{seds}
\end{figure}

\begin{figure}[htbp]
\includegraphics[width=9cm]{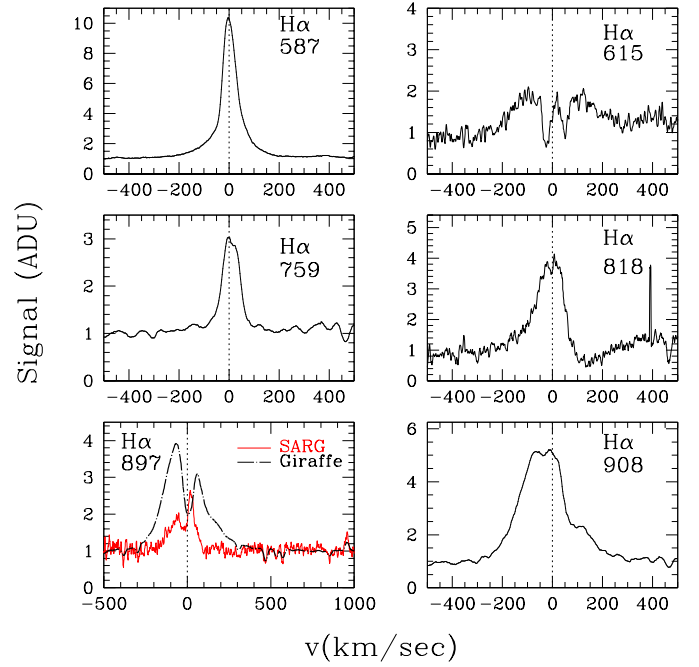}
\caption{H$\alpha$ line profiles of six stars of the sample normalized to the continuum. 
The spectra have been obtained 
either in 2004 with FLAMES/Giraffe by Sacco et al., with a spectral resolution 
R=17000, or by Rigliaco et al., in 2009 with 
SARG, with a spectral resolution R=29000, as indicated in Table~\ref{AccPROP}.}
\label{profile}
\end{figure}

In this section, we show SEDs (Fig.~\ref{seds}) and \Ha\ profiles (Fig.~\ref{profile}) 
of a subset of  two TD and four EV stars observed with SARG@TNG (Sect.~2.4) and Giraffe (Sacco et al.~2008). 
A summary of the \Ha\ properties is given in  Table~\ref{AccPROP}, 
where 10\% \Ha\ represents the width of \Ha\ at 10\% of 
the line's peak intensity, and pEW is the pseudo-equivalent width. 
Below we will briefly comment on each object. 

SO587. This EV disk was extensively studied by Rigliaco et al.~(2009).
It exhibits modest excess emission (well below the lower quartiles
of the distributions) in the IRAC bands and at 24 $\mu$m. 
It shows a symmetric and narrow
 \Ha\ emission profile with the peak close to the line center 
(type I profile following the classification 
of Reipurth et al.~(1996)). 
Based on the available U-band photometry (Wolk~1996),  the narrow \Ha\ and
the strength and profiles of the [\ion{S}{ii}] and [\ion{N}{ii}] forbidden 
lines, Rigliaco et al. (2009) 
suggested that the disk was being
photoevaporated and that the forbidden lines were
coming from the photoevaporation flow,
possibly driven and certainly illuminated by the star \sori. 
A crucial ingredient of this model was the high ratio between the mass-loss
and the mass-accretion rate.
Our results confirm this interpretation.
Although relatively bright in U, the object does not have
a measurable U-band excess (i.e., it lies inside the photospheric strip
of Fig.~\ref{calib}) and we 
estimate an upper limit to $\log$ \Macc\ of $-9.45$ \Msun/y,
consistent with the value $-9.52$ \Msun/y derived in Rigliaco et al. (2009).

SO615. This is a relatively massive, luminous star. Its SED is typical
of a flat disk up to 24 $\mu$m. We cannot measure \Macc\ from the U-band, which
is already saturated by the photospheric emission alone, but we can estimate a value 
of $\log$ \Macc\ =$-8.25$ \Msun/y from the U=15.96 mag 
measurement of Wolk  (1996). 
The \Ha\ has a complex profile, with broad wings, deep redshifted and
blueshifted absorption and a 
narrow, slightly redshifted emission in the  center. 
Similar profiles are observed for higher numbers 
of the Balmer series (``YY Orionis like profiles''; Walker, 1972) and are
 associated with extensive infall and outflow rates, consistent with the 
very high value of \Macc.
 Unfortunately, the SARG SO615 spectrum is rather
noisy.

SO759. Classified as EV disk star by Hernandez et al.~(2007).
 The SED has negligible excess up to 8 $\mu$m, but
a significant one at 24 $\mu$m, not very 
different from the SED of some TD objects. 
The \Ha\ is rather narrow and symmetric, with 
a moderate red/blue asymmetry
(type I profile, Reipurt et al.~(1996)).
The 10\% \Ha\ width of $157\pm12$ km s$^{-1}$ and 
the upper limit $\log$ \Macc\ $<-9.67$ \Msun/y
is not stringent for an object of 0.3 \Msun, but, combined with the \Ha\
properties, it suggests that this is a low accretor (if any).

SO818. Classified as TD. This star shows significant excess emission,
in the higher quartile at 3.6, 4.5, 5.8
and at 24 $\mu$m. We measure \Macc=$-9.45$ \Msun/y, typical of optically thick disks. 
The \Ha\ is broad (10\% width of $332\pm25$ km s$^{-1}$) and shows an inverse P-Cygni profile, with the emission  
peak  at the line center position. The red-shifted absorption 
goes below the continuum 
(type IVR), confirming the evidence of a high accretion rate. 

SO897. Classified as TD; the IRAC excess emission is clearly detected, but lower than the \sori\
medians, while the  24 $\mu$m excess is strong. We measure a lower limit to \Macc ($>-9.13$ \Msun/y), but the star is clearly accreting.
We have two measurements of the \Ha\ profile, one with  Giraffe 
in October 2004 and one with SARG  acquired in January 2009. Within these two epochs the maxima change in position 
and strength, with the primary one blue-shifted in 2004 and red-shifted in 2009, and the 
secondary one with opposite behaviour. The wavelength separation of the blue and red emission 
peaks decreases from 2004 and 2009, while the central reversal seems to be at the line 
center in the first epoch and then slightly blue-shifted in the second epoch. The profile 
changes from IIR to IIB, following the Reipurt et al. (1996) scheme, where these 
types are characterized by secondary peaks exceeding half the strength of the primary peaks, 
as we observe. 
This is a common phenomenon in accreting T Tauri stars, probably due
to the interplay of variable accretion and mass-loss.

SO908. EV disk, with  excess emission within the lower quartiles at all
 IRAC wavelengths, and significant excess at 24 $\mu$m.
We measure \Macc=-9.37 \Msun/y, typical of optically thick disks.
The \Ha\ is broad and asymmetric, with 
less emission in the red than in the blue.
This type of profile (IIIR) is the less frequent
in the scheme classification 
of Reipurt et al. (1996). Following the radiative transfer models developed by 
Kurosawa et al. (2006), this profile morphology requires some obscuration  
by the dusty disk, i.e. a high inclination, explaining the rarity of the profile. A highly inclined disk is consistent with the SED properties.

\begin{table*}[t]
\centering
\caption{Accretion properties of the stars shown in Fig.~\ref{profile}.
$^a$Stars names, following the nomenclature of Hernandez et al.~ 2007;
$^b$values of the mass accretion rates derived from the U-band excess emission;
$^c$width of the H$\alpha$ line at 10\% of the peak;
$^d$values of the \Macc\ obtained using the empirical relation which involves the H$\alpha$ width at 10\% of the peak (Natta et al.~2004);
$^e$pseudo-equivalent width of the H$\alpha$ line.
}
\begin{tabular}{c c c c c c}
\hline \hline
 Object$^a$ & $\log$ \Macc\ (\Msun/y)$^b$ & 10\% H$\alpha$ $^c$ & $\log$ \Macc\ (\Msun/y)$^d$& pEW$^e$ & Instrument \\
        &  (U-band)  & (km s$^{-1}$)  & (10\% H$\alpha$) & (\AA)    &  \\
\hline
 SO587 & $<$ -9.45 &  191$\pm$10 & - & -16.48$\pm$0.76 & Giraffe \\ 
 SO615 & sat    &  397$\pm$50 & -9.04 & -4.4$\pm$0.4 & Sarg \\
 SO759 & $<$ -9.67 &  157$\pm$12 & - & -3.39$\pm$0.22 & Giraffe\\  
 SO818 & -9.45  &  332$\pm$25 & -9.67 &  -6.5$\pm$0.7 & Sarg\\
 SO897 & $>$ -9.13 &  235$\pm$19 & -10.61 & -4.21$\pm$2 & Sarg  \\  
 SO897 & - &  503$\pm$57 & -8.0 & -13.19$\pm$1.38 & Giraffe \\  
 SO908 & -9.37  &  401$\pm$26 & -9.0 & -19.65$\pm$0.19 & Giraffe\\

\hline
\end{tabular}
\label{AccPROP}
\end{table*}


\longtabL{2}{
\begin{landscape}
\begin{longtable}{c c c c c c c c c c c c c c c c c}
\caption{U-band and collected literature photometry. The U-band magnitudes have been obtained with FORS1@VLT, 
the optical photometry is from Sherry et al.~(2004), Kenyon et al.~(2005), Zapatero-Osorio et al.~(2002), 
B\'ejar et al.~(2001) and Wolk~(1996).   
JHK magnitudes are from the Two Micron All Sky Survey (2MASS) (Cutri et al.~2003). 
The magnitudes in the four channels of the Infrared Array Camera (IRAC; 3.8-8.0 $\mu$m) 
and the first channel of the Multiband Imaging Photometer for {\it{Spitzer}} (MIPS; 24 $\mu$m) 
are  from Hernandez et al.~(2007). 
}\\

\hline\hline

Name & RA & DEC & class & U & B & V & R & I & J & H & K & 3.6 \um & 4.5 \um & 5.8 \um & 8.0 \um & 24.0 \um \\
	& ($\circ$) & ($\circ$) & & (mag) & (mag) & (mag) & (mag) & (mag) & (mag) & (mag) & (mag) & (mag) & (mag) & (mag) & (mag) & (mag) \\
\hline

\hline
\endfirsthead
\caption{continued.}\\
\hline\hline
\endhead
\endfoot

&&&&&&&&&&&&&&&&\\
SO209 &  84.46288 &  -2.43542 & III & $>$23.0 & ... & ... & ... & 17.135 & 14.921 & 14.374 & 13.954 & 13.64 & 13.53 & 13.68 & 13.38 & ... \\
SO251 &  84.47960 &  -2.46006 & III & 20.82 & 20.55 & 18.49 & 17.18 & 15.38 & 13.6 & 12.96 & 12.74 & 12.45 & 12.36 & 12.3 & 12.46 & ... \\ 
SO299 &  84.50394 &  -2.43548 & TD &  18.89 & 18.08 & 16.71 & 15.57 & 14.27 & 12.82 & 12.15 & 11.93 & 11.68 & 11.76 & 11.66 & 11.33 & 6.76 \\ 
SO302 &  84.50689 &  -2.43130 & III & 19.82 & 18.71 & 17.21 & 16.01 & 14.54 & 13.03 & 12.32 & 12.07 & 11.86 & 11.83 & 11.78 & 11.8 & ... \\ 
SO320 &  84.51746 &  -2.31825 & III & 17.5 & ... & 14.49 & ... & ... & 11.727 & 11.022 & 10.826 & 10.72 & 10.77 & 10.76 & 10.7 & ... \\ 
SO338 &  84.52699 &  -2.48038 & III & $<$17 & 11.807 & 11.429 & ... & 10.741 & 10.088 & 9.829 & 9.75 & 9.69 & 9.67 & 9.66 & 9.68 & 9.62 \\ 
SO341 &  84.52801 &  -2.50627 & II & 18.07 & 16.46 & 15.08 & 14.08 & 13.14 & 11.76 & 10.92 & 10.54 & 9.73 & 9.47 & 9.2 & 8.43 & 5.4 \\ 
SO352 &  84.53263 &  -2.52533 & III & $<$17 & ... & ... & ... & ... & 10.566 & 9.93 & 9.769 & 9.63 & 9.65 & 9.58 & 9.56 & 9.54 \\ 
SO366 &  84.53733 &  -2.33634 & III & 20.31 & 19.15 & 17.66 & 16.41 & 14.85 & 13.25 & 12.6 & 12.31 & 12.1 & 12.07 & 12.01 & 12.05 & ... \\
SO397 &  84.55490 &  -2.43573 & II & 18.69 & 18.41 & 16.92 & 15.67 & 14.1 & 12.48 & 11.82 & 11.55 & 11.16 & 10.96 & 10.66 & 9.98 & 7.68 \\ 
SO426 &  84.56700 &  -2.63466 & III & 20.75 & ... & 18.3 & 16.81 & 15.2 & 13.58 & 12.88 & 12.61 & 12.38 & 12.28 & 12.27 & 12.25 & ... \\ 
SO432 &  84.57150 &  -2.37377 & III & 21.92 & ... & 19.08 & 17.3 & 15.25 & 13.61 & 13.03 & 12.74 & 12.33 & 12.26 & 12.21 & 12.17 & ... \\ 
SO435 &  84.57401 &  -2.68056 & II & 20.44 & 19.72 & 18.14 & 16.81 & 14.98 & 13.2 & 12.58 & 12.24 & 11.77 & 11.52 & 11.32 & 10.67 & 7.57 \\ 
SO440 &  84.57490 &  -2.35288 & III & 17.07 & ... & 14.43 & 13.32 & 12.53 & 11.77 & 11.09 & 10.88 & 10.81 & 10.89 & 10.86 & 10.78 & ... \\ 
SO444 &  84.57595 &  -2.80397 & EV & 19.23 & 18.14 & 16.68 & 15.54 & 14.15 & 12.76 & 12.02 & 11.8 & 11.62 & 11.57 & 11.51 & 11.29 & ... \\ 
SO460 &  84.58411 &  -2.63373 & III & 19.81 & ... & 17.49 & 16.02 & 14.33 & 12.58 & 11.86 & 11.61 & 11.32 & 11.24 & 11.19 & 11.25 & ... \\ 
SO462 &  84.58536 &  -2.56910 & II & 19.32 & 18.94 & 17.29 & 16.07 & 14.36 & 12.65 & 11.92 & 11.65 & 11.03 & 10.63 & 10.14 & 8.97 & 5.9 \\ 
SO465 &  84.58689 &  -2.77032 & III & 23.5 & ... & ... & ... & 17.41 & 15.186 & 14.572 & 14.162 & 13.81 & 13.66 & 13.53 & 13.64 & ... \\ 
SO469 &  84.58898 &  -2.56002 & III & $>$23 & ... & ... & ... & 17.612 & 15.355 & 14.79 & 14.494 & 14 & 13.85 & 13.73 & 13.82 & ... \\ 
SO475 &  84.59063 &  -2.36375 & III & $<$17 & ... & 13.36 & 12.34 & 11.62 & 10.972 & 10.365 & 10.213 & 10.08 & 10.16 & 10.15 & 10.09 & 10.11 \\ 
SO482 &  84.59605 &  -2.61371 & II & 19.94 & ... & 18.66 & 17.14 & 15.67 & 13.8 & 13.17 & 12.78 & 12.22 & 11.8 & 11.3 & 10.76 & ... \\ 
SO485 &  84.59714 &  -2.42627 & II & 18.54 & 19.6 & 18.4 & 17.22 & 15.69 & 13.69 & 12.93 & 12.42 & 11.9 & 11.58 & 11.22 & 10.53 & 7.83 \\ 
SO489 &  84.59799 &  -2.69212 & III & 20.74 & ... & 18.26 & 16.84 & 15.08 & 13.29 & 12.74 & 12.4 & 12.16 & 12.04 & 11.98 & 11.99 & ... \\ 
SO490 &  84.59823 &  -2.34652 & II & 19.25 & 19.81 & 18.64 & 17.14 & 15.32 & 13.41 & 12.8 & 12.49 & 11.97 & 11.66 & 11.24 & 10.39 & 7.64 \\ 
SO500 &  84.60588 &  -2.71143 & II & 20.73 & ... & ... & ... & 17.3 & 14.88 & 14.16 & 13.57 & 12.75 & 12.37 & 12 & 11.37 & 8.59 \\ 
SO502 &  84.60695 &  -2.52263 & III & 22.82 & ... & ... & ... & 16.627 & 14.666 & 14.07 & 13.837 & 13.38 & 13.32 & 13.31 & 13.31 & ... \\ 
SO509 &  84.60924 &  -2.67810 & III & 23.37 & ... & ... & ... & 17.283 & 14.909 & 14.281 & 13.918 & 13.55 & 13.48 & 13.45 & 13.46 & ... \\ 
SO514 &  84.61174 &  -2.64612 & II & 22.11 & ... & ... & 18.03 & 16.06 & 14.11 & 13.48 & 13.21 & 12.57 & 12.27 & 11.99 & 11.29 & 8.3 \\ 
SO518 &  84.61349 &  -2.75263 & II & $<$17 & 16.1 & 14.192 & 13.64 & 12.85 & 11.955 & 10.792 & 9.944 & 8.75 & 8.36 & 7.98 & 7.18 & 4.62 \\ 
SO520 &  84.61455 &  -2.58446 & II & 19.23 & ... & 17.09 & 15.84 & 14.38 & 12.83 & 12.11 & 11.86 & 11.43 & 11.16 & 10.68 & 9.73 & 6.62 \\ 
SO521 &  84.61462 &  -2.72567 & III & $<$16.5 & 10.9 & 10.8 & 11.2 & 10.396 & 10.176 & 10.099 & 10.103 & 10.04 & 10.04 & 10.05 & 10.05 & 9.61 \\ 
SO525 &  84.61552 &  -2.71691 & III & 18.81 & ... & 16.29 & 15 & 13.67 & 12.19 & 11.45 & 11.27 & 11.11 & 11.06 & 11.01 & 11.02 & ... \\ 
SO537 &  84.62065 &  -2.81309 & II & $>$23.27 & ... & ... & ... & 16.952 & 14.823 & 14.277 & 13.877 & 13.1 & 12.82 & 12.48 & 11.93 & 9.42 \\ 
SO539 &  84.62125 &  -2.60070 & III & 19.01 & ... & 16.34 & 15.22 & 14.04 & 12.63 & 11.89 & 11.69 & 11.53 & 11.52 & 11.49 & 11.46 & ... \\ 
SO540 &  84.62143 &  -2.27104 & II & $<$17 & 15.4 & 14.49 & 13.2 & 12.954 & 11.697 & 11.021 & 10.759 & 10.51 & 10.41 & 10.26 & 9.42 & 4.99 \\ 
SO545 &  84.62334 &  -2.42061 & III & 23.32 & ... & ... & ... & 18.73 & 17.046 & 14.84 & 14.29 & 13.96 & 13.56 & 13.49 & 13.52 & 13.51 \\ 
SO550 &  84.62529 &  -2.35547 & III & 17.19 & ... & 14.81 & 13.73 & 12.96 & 12.2 & 11.55 & 11.39 & 11.28 & 11.35 & 11.33 & 11.27 & ... \\ 
SO557 &  84.62903 &  -2.56769 & III & 22.22 & ... & ... & ... & ... & 14.92 & 14.26 & 14.009 & 13.71 & 13.65 & 13.68 & 13.42 & ... \\ 
SO562 &  84.63079 &  -2.60936 & II & 16.98 & 16.64 & 16.3 & 15.09 & 13.78 & 12.174 & 11.473 & 10.986 & 10.32 & 9.88 & 9.51 & 8.5 & 5.57 \\ 
SO563 &  84.63153 &  -2.58742 & II & 17.84 & ... & 15.64 & 14.6 & 13.49 & 11.52 & 10.71 & 10.35 & 9.71 & 9.3 & 8.9 & 8.3 & 6.05 \\ 
SO568 &  84.63512 &  -2.49917 & III & $>$23 & ... & ... & ... & 17.578 & 15.439 & 14.84 & 14.44 & 14.09 & 13.96 & 13.9 & 13.96 & ... \\ 
SO572 &  84.63676 &  -2.59419 & III & 17.04 & ... & ... & 13.71 & 12.739 & 11.544 & 10.896 & 10.73 & 10.57 & 10.54 & 10.57 & 10.53 & ... \\ 
SO576 &  84.63754 &  -2.65777 & III & 21.58 & ... & ... & ... & ... & 14.59 & 14.02 & 13.7 & 13.44 & 13.4 & 13.18 & 13.3 & ... \\ 
SO582 &  84.63890 &  -2.60486 &  III & 18.64 & ... & 15.87 & 15.51 & 13.39 & 12.052 & 11.295 & 11.107 & 10.95 & 10.94 & 10.91 & 10.91 & ... \\ 
SO583 &  84.64026 &  -2.73726 & II & $<$16.5 & 13.5 & 13.6 & ... & 11.509 & 10.131 & 9.28 & 8.666 & 7.68 & 7.37 & 7.02 & 6.05 & 2.95 \\
SO587 &  84.64184 &  -2.61037 & EV & 18.83 & ... & 16.39 & 15.24 & 13.72 & 11.98 & 11.33 & 11.08 & 10.73 & 10.56 & 10.43 & 9.86 & 6.92 \\ 
SO590 &  84.64260 &  -2.57110 & III & $<$16.5 & 8.543 & 8.556 & ... & 8.57 & 8.779 & 8.82 & 8.79 & 8.84 & 8.8 & 8.76 & 8.8 & 8.65 \\ 
SO592 &  84.64287 &  -2.58333 & III & $<$17 & ... & ... & ... & ... & 11.22 & 10.56 & 10.35 & 10.26 & 10.36 & 10.22 & 10.23 & ... \\ 
SO598 &  84.64410 &  -2.68572 & II & 19.56 & ... & 17.76 & 16.18 & 14.72 & 13.1 & 12.45 & 12.12 & 11.77 & 11.62 & 11.48 & 10.88 & 8.56 \\ 
SO601 &  84.64494 &  -2.57098 & III & $<$17 & 8.395 & 8.388 & ... & 8.4 & 8.346 & 8.38 & 8.374 & 8.32 & 8.36 & 8.34 & 8.36 & 8.36 \\ 
SO602 &  84.64509 &  -2.54782 &  III & $<$17 & ... & ... & ... & 10.743 & 9.89 & 9.302 & 9.211 & 9.15 & 9.17 & 9.08 & 9.1 & 8.86 \\ 
SO609 &  84.64726 &  -2.42280 & III & $>$23.0 & ... & ... & ... & 16.878 & 14.652 & 14.056 & 13.764 & 13.34 & 13.23 & 13.25 & 13.19 & ... \\ 
SO611 &  84.64769 &  -2.53098 & III & 16.67 & ... & ... & 13.24 & 12.501 & 11.303 & 10.627 & 10.46 & 10.34 & 10.33 & 10.28 & 10.23 & ... \\ 
SO615 &  84.64941 &  -2.73083 & EV & $<$17 & 12.9 & ... & 13.2 & 12.438 & 10.445 & 9.726 & 9.311 & 8.6 & 8.26 & 7.99 & 7.73 & 5.99 \\ 
SO616 &  84.64942 &  -2.51199 & III & 16.52 & 17 & 13.701 & 13.28 & 12.48 & 11.245 & 10.598 & 10.424 & 10.32 & 10.29 & 10.24 & 10.21 & ... \\ 
SO620 &  84.65217 &  -2.55346 & III & $<$16.5 & 7.77 & 7.88 & ... & 7.99 & 8.1 & 8.18 & 8.202 & 8.22 & 8.28 & 8.2 & 8.26 & 7.75 \\ 
SO621 &  84.65282 &  -2.73709 & III & 20.09 & ... & 17.65 & 16.08 & 14.32 & 12.56 & 11.91 & 11.62 & 11.37 & 11.3 & 11.26 & 11.23 & ... \\ 
SO628 &  84.65598 &  -2.83985 & III & 20.28 & 19.33 & 17.85 & 16.43 & 14.66 & 12.81 & 12.18 & 11.92 & 11.62 & 11.55 & 11.55 & 11.51 & ... \\ 
SO631 &  84.65773 &  -2.34434 & III & $>$23 & ... & ... & 19.55 & 18 & 15.6 & 14.92 & 14.64 & 14.14 & 14.04 & 14.01 & 14.36 & ... \\ 
SO638 &  84.66028 &  -2.58193 & EV & $<$17 & ... & ... & 11.73 & ... & 9.907 & 9.28 & 9.119 & 8.96 & 9.07 & 8.86 & 8.76 & 6.34 \\ 
SO641 &  84.66072 &  -2.69883 & III & 21.65 & ... & ... & 18.28 & 16.36 & 14.56 & 13.97 & 13.65 & 13.31 & 13.25 & 13.16 & 13.2 & ... \\ 
SO644 &  84.66195 &  -2.46712 & III & 22.88 & 20.3 & ... & 18.77 & 17.07 & 15.27 & 14.73 & 14.43 & 14.11 & 14.04 & 13.99 & 13.91 & ... \\ 
SO646 &  84.66254 &  -2.75888 & II & 17.66 & ... & 17.26 & 15.92 & 14.58 & 12.91 & 12.2 & 11.89 & 11.19 & 10.86 & 10.5 & 9.92 & 7.98 \\ 
SO648 &  84.66335 &  -2.88562 & III & 18.16 & ... & 15.59 & ... & ... & 12.697 & 12.04 & 11.87 & 11.79 & 11.8 & 11.76 & 11.74 & ... \\ 
SO655 &  84.66545 &  -2.67210 & III & 20.68 & ... & ... & ... & ... & 13.75 & 13.1 & 12.88 & 12.55 & 12.48 & 12.48 & 12.49 & ... \\ 
SO657 &  84.66563 &  -2.53893 & II & $>$23 & ... & ... & ... & 17.556 & 14.89 & 14.284 & 13.942 & 13.18 & 12.82 & 12.44 & 11.7 & 8.61 \\ 
SO658 &  84.66696 &  -2.84362 & III & 21.08 & 20.15 & 18.66 & 17.26 & 15.48 & 13.67 & 13.08 & 12.8 & 12.52 & 12.46 & 12.43 & 12.42 & ... \\ 
SO662 &  84.66770 &  -2.50515 & II & 17.22 & 15.98 & 14.27 & 13.74 & 12.89 & 11.51 & 10.76 & 10.4 & 9.84 & 9.42 & 8.88 & 8.08 & 5.51 \\ 
SO663 &  84.66877 &  -2.55761 & II & 20.68 & ... & 17.64 & 16.19 & 14.54 & 12.82 & 12.13 & 11.87 & 11.45 & 11.22 & 10.94 & 10.34 & 7.3 \\ 
SO674 &  84.67322 &  -2.50798 & II & 19.52 & ... & 17.04 & 15.84 & 14.39 & 12.84 & 12.14 & 11.93 & 11.69 & 11.56 & 11.29 & 10.55 & 7.44 \\ 
SO692 &  84.68226 &  -2.87850 & III & 20.19 & 19.59 & 18.07 & 16.63 & 14.86 & 13.01 & 12.39 & 12.1 & 11.81 & 11.72 & 11.66 & 11.68 & ... \\ 
SO697 &  84.68423 &  -2.67209 & II & 16.71 & ... & ... & ... & 12.51 & 11.36 & 10.69 & 10.44 & 9.83 & 9.41 & 8.90 & 7.81 & 5.20 \\ 
SO700 &  84.68530 &  -2.67708 & EV & $>$23 & ... & ... & 19.424 & 17.297 & 14.802 & 14.213 & 13.935 & 13.33 & 13.25 & 12.99 & 12.72 & ... \\ 
SO701 &  84.68532 &  -2.67507 & III & 20.28 & ... & ... & ... & 14.99 & 13.37 & 12.72 & 12.5 & 12.24 & 12.22 & 12.24 & 12.23 & ... \\ 
SO710 &  84.68900 &  -2.69983 & II & 17.63 & 15.96 & 15.73 & 14.62 & 13.46 & 11.99 & 11.33 & 11.04 & 10.55 & 10.28 & 9.95 & 8.91 & 6.15 \\ 
SO714 &  84.69149 &  -2.75640 & III & 20.97 & ... & 19.15 & ... & 15.49 & 13.56 & 12.96 & 12.69 & 12.34 & 12.27 & 12.24 & 12.26 & ... \\ 
SO728 &  84.69813 &  -2.45337 & EV & 19.69 & ... & ... & ... & 14.05 & 12.14 & 11.5 & 11.27 & 10.91 & 10.86 & 10.74 & 10.57 & ... \\ 
SO730 &  84.69852 &  -2.51034 & III & 21.47 & ... & ... & ... & 15.39 & 13.45 & 12.85 & 12.59 & 12.22 & 12.26 & 12.11 & 12.15 & ... \\ 
SO736 &  84.70019 &  -2.45398 & II & $<$17 & 14.4 & 13.34 & ... & 11.36 & 10.156 & 9.463 & 9.187 & 8.62 & 8.34 & 8.03 & 7.49 & 5.53 \\ 
SO738 &  84.70032 &  -2.48150 & II & 22.4 & ... & ... & ... & 16.44 & 14.47 & 13.84 & 13.44 & 13 & 12.72 & 12.37 & 11.55 & 8.58 \\ 
SO739 &  84.70071 &  -2.73354 & II & 22.34 & ... & ... & ... & 16.22 & 14.07 & 13.47 & 13.15 & 12.58 & 12.29 & 11.9 & 11.19 & 8.61 \\ 
SO747 &  84.70481 &  -2.63955 & III & $<$17 & ... & 15.05 & 13.99 & 12.89 & 11.4 & 10.66 & 10.51 & 10.29 & 10.29 & 10.21 & 10.23 & ... \\ 
SO748 &  84.70501 &  -2.69025 & III & 18.45 & ... & 15.98 & 14.73 & 13.23 & 11.7 & 11.01 & 10.7 & 10.57 & 10.53 & 10.54 & 10.49 & ... \\ 
SO750 &  84.70535 &  -2.39936 & II & 21.08 & ... & ... & 17.67 & 16.573 & 14.362 & 13.699 & 13.2 & 12.79 & 12.47 & 12.01 & 11.57 & 9.5 \\ 
SO757 &  84.70797 &  -2.68965 & III & 19.33 & ... & 16.97 & 15.58 & 14.21 & 12.77 & 12.03 & 11.8 & 11.58 & 11.56 & 11.55 & 11.53 & ... \\ 
SO759 &  84.70995 &  -2.44663 & EV & 19.35 & ... & 16.79 & 15.58 & 14.1 & 12.5 & 11.84 & 11.54 & 11.37 & 11.23 & 11.09 & 10.85 & 8.44 \\ 
SO762 &  84.71085 &  -2.71188 & II & 21.77 & ... & 19.29 & ... & 15.89 & 13.84 & 13.25 & 12.96 & 12.44 & 12.15 & 11.71 & 11.03 & 8.16 \\ 
SO767 &  84.71257 &  -2.82059 & III & 22.81 & ... & ... & ... & 16.9 & 15.04 & 14.42 & 14.16 & 13.79 & 13.71 & 13.56 & 13.74 & ... \\ 
SO774 &  84.71678 &  -2.77881 & II & 16.96 & 16.2 & 14.301 & 13.76 & 12.72 & 11.518 & 10.774 & 10.421 & 10.18 & 9.95 & 9.57 & 8.42 & 5.02 \\ 
SO787 &  84.72228 &  -2.55633 & III & $<$17 & ... & ... & 12.42 & 11.47 & 10.607 & 9.919 & 9.734 & 9.59 & 9.56 & 9.48 & 9.5 & 9.37 \\ 
SO791 &  84.72552 &  -2.82499 & III & $<$17 & 15 & 13.164 & 13.6 & 11.73 & 10.829 & 10.31 & 10.126 & 10.16 & 10.14 & 9.98 & 10.11 & 9.97 \\ 
SO795 &  84.72658 &  -2.66761 & III & 21.91 & ... & 19.11 & ... & ... & 14.311 & 13.698 & 13.381 & 13.04 & 13.03 & 12.97 & 12.92 & ... \\ 
SO797 &  84.72879 &  -2.48283 & III & 21.05 & 19.84 & 18.61 & 17.26 & 15.5 & 13.8 & 13.2 & 12.87 & 12.6 & 12.55 & 12.47 & 12.48 & ... \\ 
SO818 &  84.74306 &  -2.26953 & TD & 17.53 & 17.04 & 15.61 & 14.62 & 13.64 & 12.34 & 11.56 & 11.28 & 10.51 & 10.37 & 10.17 & 10.06 & 6.39 \\ 
SO827 &  84.74685 &  -2.56428 & II & 18.96 & 18.24 & 17.14 & 16.09 & 14.83 & 12.888 & 11.979 & 11.401 & 10.87 & 10.47 & 10.09 & 9.27 & 6.42 \\ 
SO841 &  84.75486 &  -2.61078 & III & 20.51 & ... & 18.04 & ... & 15.49 & 13.52 & 12.9 & 12.61 & 12.46 & 12.36 & 12.43 & 12.3 & ... \\ 
SO844 &  84.75575 &  -2.30771 & II & 16.97 & 16.28 & 15.01 & 14.02 & 12.99 & 11.73 & 10.84 & 10.34 & 8.93 & 8.52 & 8.2 & 7.53 & 4.55 \\ 
SO847 &  84.75634 &  -2.64902 & III & $<$16 & 7.97 & 8.06 & ... & 8.21 & 8.131 & 8.105 & 8.093 & 8.07 & 8.06 & 8.06 & 8.06 & 7.73 \\ 
SO848 &  84.75808 &  -2.58412 & II & 19.63 & ... & 19.08 & ... & 16.38 & 14.45 & 13.38 & 12.61 & 11.62 & 11.04 & 10.6 & 9.84 & 6.45 \\ 
SO855 &  84.76159 &  -2.49887 & III & 19.6 & 18.42 & 16.91 & 15.72 & 14.21 & 12.61 & 12 & 11.69 & 11.47 & 11.39 & 11.35 & 11.37 & ... \\ 
SO859 &  84.76255 &  -2.69087 & II & 18.62 & 17.47 & 16.46 & 15.4 & 14.11 & 12.44 & 11.61 & 11.16 & 10.34 & 9.98 & 9.7 & 9.12 & 6.47 \\ 
SO865 &  84.76500 &  -2.77413 & II & 18.54 & 18.04 & 16.94 & 15.82 & 14.37 & 12.84 & 12.12 & 11.86 & 11.47 & 11.22 & 10.9 & 10.12 & 6.83 \\ 
SO866 &  84.76612 &  -2.33562 & II & 20.26 & 20.09 & 18.79 & 17.38 & 15.62 & 13.83 & 13.16 & 12.88 & 12.4 & 12.08 & 11.66 & 10.93 & 8.34 \\ 
SO869 &  84.76859 &  -2.87525 & III & 17.17 & ... & 14.54 & ... & ... & 11.66 & 10.988 & 10.814 & 10.7 & 10.68 & 10.59 & 10.63 & ... \\ 
SO877 &  84.77193 &  -2.55015 & III & 20.4 & 19.29 & 17.75 & 16.57 & 15.02 & 13.39 & 12.72 & 12.46 & 12.21 & 12.25 & 12.12 & 12.14 & ... \\ 
SO879 &  84.77262 &  -2.54176 & III & 17.06 & 14.7 & 14.44 & 13.53 & 12.83 & 11.55 & 10.86 & 10.67 & 10.58 & 10.66 & 10.59 & 10.55 & ... \\ 
SO896 &  84.78171 &  -2.47320 & III & 19.49 & 18.4 & 16.93 & 15.82 & 14.37 & 12.88 & 12.14 & 11.96 & 11.74 & 11.69 & 11.66 & 11.67 & ... \\ 
SO897 &  84.78176 &  -2.54423 & TD & $<$17 & 15.5 & ... & 13.44 & 12.921 & 11.298 & 10.573 & 10.26 & 9.75 & 9.65 & 9.58 & 9.31 & 4.98 \\
SO902 &  84.78430 &  -2.54124 & III & 21.67 & 19.19 & 18.83 & 17.55 & 15.76 & 13.8 & 13.25 & 12.92 & 12.62 & 12.55 & 12.59 & 12.63 & ... \\ 
SO905 &  84.78563 &  -2.86296 & EV & 17.95 & 16.64 & 15.23 & 14.21 & 13.23 & 11.95 & 11.2 & 11.03 & 10.83 & 10.73 & 10.65 & 10.36 & 8.89 \\ 
SO908 &  84.78665 &  -2.51992 & EV & 19.03 & 19.36 & 17.91 & 16.6 & 14.97 & 13.04 & 12.16 & 11.7 & 11.07 & 10.91 & 10.81 & 10.35 & 6.72 \\ 
SO911 &  84.78742 &  -2.66611 & III & 22.76 & ... & ... & ... & 17.37 & 14.66 & 14.13 & 13.74 & 13.26 & 13.16 & 13.26 & 13.12 & ... \\ 
SO917 &  84.79183 &  -2.46993 & EV & 21.52 & ... & ... & 17.68 & 16.05 & 14.6 & 13.99 & 13.78 & 13.56 & 13.37 & 13.4 & 13.09 & ... \\ 
SO924 &  84.79681 &  -2.43031 & III & $<$16.5 & ... & 15.32 & ... & ... & 10.963 & 10.593 & 10.475 & 9.62 & 9.03 & 8.35 & 7.18 & 4.64 \\ 
SO925 &  84.79756 &  -2.55911 & III & 22.57 & ... & ... & ... & 16.54 & 14.45 & 13.93 & 13.57 & 13.21 & 13.12 & 13.02 & 13.03 & ... \\ 
SO927 &  84.79804 &  -2.51851 & II & 18.14 & 16.31 & 15.32 & 14.4 & 13.45 & 11.99 & 11.19 & 10.73 & 9.62 & 9.03 & 8.35 & 7.18 & 4.64 \\ 
SO929 &  84.79854 &  -2.60081 & III & 17.27 & ... & 14.79 & 13.88 & 12.93 & 11.65 & 10.97 & 10.75 & 10.61 & 10.61 & 10.57 & 10.51 & ... \\ 
SO931 &  84.79945 &  -2.46140 & III & 20.63 & 19.64 & 18.03 & 16.84 & 15.2 & 13.61 & 12.98 & 12.65 & 12.52 & 12.39 & 12.39 & 12.29 & ... \\ 
SO933 &  84.80144 &  -2.50184 & III & 20.68 & 20.44 & 18.17 & 16.69 & 14.69 & 12.61 & 12.06 & 11.73 & 11.37 & 11.28 & 11.22 & 11.23 & ... \\ 
SO936 &  84.80457 &  -2.63085 & II & $>$23.2 & ... & ... & ... & 17.793 & 15.24 & 14.747 & 14.311 & 13.83 & 13.53 & 13.29 & 12.61 & ... \\ 
SO940 &  84.80616 &  -2.62760 & III & 20.91 & ... & 18.33 & ... & 15.65 & 13.41 & 12.77 & 12.5 & 12.21 & 12.15 & 12.1 & 12.04 & ... \\ 
SO946 &  84.81039 &  -2.47597 & III & 20.17 & 19.03 & 17.57 & 16.37 & 14.85 & 13.34 & 12.65 & 12.34 & 12.14 & 12.08 & 12.05 & 12.02 & ... \\ 
SO957 &  84.81302 &  -2.67986 & III & 23.26 & ... & ... & ... & 17.279 & 14.669 & 14.042 & 13.656 & 13.31 & 13.21 & 13.11 & 13.14 & ... \\ 
SO961 &  84.81527 &  -2.49919 & III & $<$17 & ... & ... & ... & ... & 9.808 & 9.498 & 9.398 & 9.4 & 9.37 & 9.33 & 9.32 & 9.38 \\
SO967 &  84.81602 &  -2.61414 & II & 20.05 & 18.6 & 17.6 & 16.36 & 15.21 & 13.25 & 12.54 & 12.22 & 11.56 & 11.29 & 11.09 & 10.67 & 8.34 \\ 
SO976 &  84.82092 &  -2.68808 & III & 21.27 & 19.6 & 18.6 & ... & 16.31 & 14.29 & 13.63 & 13.37 & 13.1 & 13.03 & 13.01 & 12.97 & ... \\ 
SO978 &  84.82169 &  -2.42875 & III & 19.08 & 17.98 & 16.4 & 15.33 & 14.18 & 12.9 & 12.12 & 11.93 & 11.81 & 11.71 & 11.64 & 11.64 & ... \\ 
SO981 &  84.82534 &  -2.49126 & EV & $<$17 & ... & ... & ... & 11.729 & 10.721 & 10.27 & 10.117 & 10.08 & 10.04 & 9.98 & 9.98 & 8.64 \\ 
SO984 &  84.82854 &  -2.51480 & II & 17.22 & 16 & ... & 13.42 & 12.58 & 11.4 & 10.64 & 10.34 & 9.94 & 9.53 & 9.27 & 8.52 & 5.82 \\ 
SO999 &  84.83439 &  -2.64050 & III & 21.41 & 21.12 & 18.97 & 17.57 & 15.56 & 13.61 & 13.04 & 12.78 & 12.49 & 12.32 & 12.27 & 12.26 & ... \\ 
SO1000 &  84.83535 &  -2.46025 & III & 18.55 & 17.43 & 15.93 & 14.81 & 13.52 & 12.15 & 11.43 & 11.17 & 10.98 & 10.94 & 10.89 & 10.89 & ... \\ 
SO1005 &  84.83749 &  -2.50929 & III & 21.34 & 19.7 & ... & 17.9 & 15.74 & 13.29 & 12.75 & 12.44 & 12.09 & 11.97 & 11.93 & 11.82 & ... \\ 
SO1009 &  84.84067 &  -2.73437 & EV & 17.12 & ... & ... & ... & 12.746 & 11.097 & 10.405 & 10.218 & 10.12 & 10.14 & 10.06 & 10.06 & 9.56 \\ 
SO1017 &  84.84533 &  -2.55921 & III & 19.08 & 17.95 & 16.47 & 15.4 & 14.39 & 12.83 & 12.13 & 11.87 & 11.63 & 11.63 & 11.59 & 11.61 & ... \\ 
SO1027 &  84.85155 &  -2.56708 & III & 19.23 & 18.19 & 16.62 & 15.53 & 14.28 & 12.98 & 12.27 & 12.06 & 11.81 & 11.82 & 11.71 & 11.76 & ... \\ 
SO1036 &  84.85511 &  -2.63945 & II & 17.0 & 15.88 & 14.72 & 13.78 & 12.83 & 11.31 & 10.451 & 10.002 & 9.14 & 8.85 & 8.55 & 7.95 & 4.88 \\ 
SO1037 &  84.85531 &  -2.46341 & III & 22.48 & ... & ... & 18.41 & 16.86 & 15.549 & 14.794 & 14.561 & 14.32 & 14.28 & 14.4 & 14.2 & ... \\ 
SO1043 &  84.85680 &  -2.56789 & III & 20.61 & 20.12 & 18.22 & 16.71 & 14.96 & 13.2 & 12.54 & 12.25 & 11.95 & 11.86 & 11.77 & 11.83 & ... \\ 
SO1050 &  84.85986 &  -2.47716 & II & 20.74 & 19.69 & 18.02 & 16.96 & 15.26 & 13.5 & 12.84 & 12.57 & 12.21 & 11.96 & 11.69 & 10.85 & 8.17 \\ 
SO1052 &  84.86042 &  -2.43767 & III & 19.71 & 18.55 & 17.08 & 15.98 & 14.81 & 13.4 & 12.67 & 12.46 & 12.28 & 12.24 & 12.21 & 12.16 & ... \\ 
SO1053 &  84.86047 &  -2.87098 & III & 19.86 & 18.91 & 17.33 & 16.09 & 14.58 & 13.04 & 12.32 & 12.07 & 11.92 & 11.79 & 11.72 & 11.71 & ... \\
SO1057 &  84.86163 &  -2.71621 & EV & 20.77 & 21.49 & 18.9 & 17.25 & 15.52 & 13.18 & 12.4 & 12.12 & 11.85 & 11.76 & 11.62 & 11.18 & 8.88 \\ 
SO1059 &  84.86194 &  -2.61567 & II & $>$23.0 & ... & ... & ... & ... & 15.461 & 14.84 & 14.488 & 13.79 & 13.48 & 13.1 & 12.43 & 9.54 \\ 
SO1075 &  84.87243 &  -2.45586 & II & $<$16.79 & ... & ... & ... & 14.81 & 12.84 & 12.02 & 11.46 & 10.74 & 10.28 & 9.67 & 8.79 & 5.56 \\ 
SO1083 &  84.88062 &  -2.81475 & III & 20.39 & 19.72 & 18.18 & 16.86 & 15.2 & 13.616 & 12.933 & 12.665 & 12.45 & 12.38 & 12.29 & 12.4 & ... \\ 
SO1092 &  84.88490 &  -2.46588 & III & 16.56 & ... & ... & ... & 12.682 & 11.175 & 10.498 & 10.326 & 10.21 & 10.27 & 10.23 & 10.18 & 10.12 \\ 
SO1094 &  84.88578 &  -2.66231 & III & $<$17 & 15.4 & 13.756 & 13.07 & 12.38 & 10.82 & 10.104 & 9.917 & 9.76 & 9.88 & 9.67 & 9.68 & 9.62 \\ 
SO1097 &  84.88722 &  -2.79700 & III & 17.65 & 16.5 & 15.09 & 14.09 & 13.1 & 11.82 & 11.12 & 10.91 & 10.76 & 10.76 & 10.71 & 10.7 & ... \\ 
SO1104 &  84.89077 &  -2.34445 & III & 18.51 & 17.19 & 15.74 & 14.73 & 13.64 & 12.37 & 11.6 & 11.43 & 11.31 & 11.3 & 11.22 & 11.09 & ... \\ 
SO1108 &  84.89308 &  -2.64641 & III & $>$23.0 & ... & ... & ... & 17.48 & 14.763 & 14.188 & 13.787 & 13.39 & 13.27 & 13.25 & 13.1 & ... \\ 
SO1113 &  84.89641 &  -2.79165 & III & 16.65 & ... & ... & ... & ... & 11.775 & 11.069 & 10.953 & 10.79 & 10.82 & 10.81 & 10.75 & ... \\ 
SO1129 &  84.90363 &  -2.84027 & III & $<$17 & ... & ... & ... & 11.093 & 9.68 & 8.986 & 8.808 & 8.72 & 8.8 & 8.69 & 8.66 & 8.65 \\ 
SO1133 &  84.90544 &  -2.44914 & III & $<$17 & 16.25 & 14.84 & 13.89 & 12.98 & 11.698 & 10.974 & 10.773 & 10.66 & 10.65 & 10.61 & 10.58 & ... \\ 
SO1137 &  84.90753 &  -2.47907 & III & 16.66 & ... & ... & ... & ... & 11.319 & 10.671 & 10.554 & 10.42 & 10.39 & 10.38 & 10.35 & ... \\ 
SO1151 &  84.91396 &  -2.54043 & III & 21.29 & 20.35 & 18.78 & 17.34 & 15.52 & 13.44 & 12.9 & 12.53 & 12.23 & 12.14 & 12.14 & 12.14 & ... \\ 
SO1156 &  84.91740 &  -2.34667 & II & $<$17 & 14.7 & 14.16 & ... & 12.715 & 11.495 & 10.632 & 10.029 & 9.91 & 9.74 & 9.55 & 8.93 & 4.73 \\ 
SO1162 &  84.91906 &  -2.65359 & III & $>$23.0 & ... & ... & 18.89 & 17.293 & 15.4 & 14.669 & 14.408 & 14.13 & 14.03 & 14.02 & 14.1 & ... \\ 
SO1163 &  84.91906 &  -2.42975 & III & $<$16 & 12 & 11.48 & 12.4 & 10.978 & 10.312 & 10.141 & 10.033 & 10.04 & 9.94 & 9.91 & 9.95 & ... \\ 
SO1182 &  84.92999 &  -2.54543 & II & 19.63 & 19.34 & 17.72 & 16.45 & 14.83 & 13.03 & 12.3 & 11.91 & 11.12 & 10.79 & 10.52 & 9.89 & 6.34 \\ 
SO1186 &  84.93156 &  -2.77430 & III & 19.81 & ... & 17.26 & ... & ... & 13.481 & 12.769 & 12.508 & 12.33 & 12.34 & 12.28 & 12.27 & ... \\ 
SO1193 &  84.93547 &  -2.41204 & II & 20.62 & 19.1 & ... & 18.02 & 16.3 & 14.17 & 13.54 & 13.15 & 12.63 & 12.35 & 12.04 & 11.29 & 8.14 \\ 
SO1204 &  84.94250 &  -2.67568 & III & $<$16 & 9.08 & 9.01 & ... & 8.94 & 8.876 & 8.894 & 8.8 & 8.89 & 8.86 & 8.82 & 8.85 & 8.84 \\
SO1207 &  84.94425 &  -2.44208 & III & 20.45 & 19.65 & 18.05 & 16.65 & 14.87 & 12.99 & 12.36 & 12.05 & 11.77 & 11.64 & 11.65 & 11.69 & ... \\ 
SO1215 &  84.94760 &  -2.43790 &  III & $<$17 & ... & ... & ... & 10.962 & 10.035 & 9.526 & 9.382 & 9.2 & 9.21 & 9.19 & 9.15 & 9.03 \\ 
SO1216 &  84.94880 &  -2.60644 & III & 20.47 & ... & 17.9 & 16.57 & 15.06 & 13.473 & 12.772 & 12.528 & 12.29 & 12.23 & 12.17 & 12.25 & ... \\ 
SO1217 &  84.94939 &  -2.54032 & III & $<$17 & ... & ... & ... & ... & 10.969 & 10.287 & 10.082 & 10.05 & 9.98 & 9.93 & 9.93 & ... \\ 
SO1220 &  84.95028 &  -2.76591 & III & 19.31 & 18.1 & 16.58 & 15.49 & 14.18 & 12.92 & 12.28 & 12.03 & 11.82 & 11.81 & 11.72 & 11.8 & ... \\ 
SO1230 &  84.95609 &  -2.39614 & II & 20.02 & 20.21 & 18.15 & 16.86 & 15.12 & 13.4 & 12.76 & 12.44 & 12.07 & 11.91 & 11.61 & 10.99 & 8.58 \\ 
SO1238 &  84.96074 &  -2.57051 & III & 20.97 & 19.4 & 18.43 & 17.9 & 15.52 & 13.68 & 13 & 12.73 & 12.47 & 12.5 & 12.38 & 12.39 & ... \\ 
SO1250 &  84.96875 &  -2.53406 & III & $<$16.94 & ... & ... & ... & ... & 11.506 & 10.879 & 10.658 & 10.55 & 10.52 & 10.47 & 10.47 & ... \\ 
SO1260 &  84.97346 &  -2.56191 & II & 16.79 & 16.8 & ... & ... & 14.44 & 12.825 & 12.064 & 11.59 & 10.87 & 10.39 & 10.13 & 9.22 & 6.36 \\ 
SO1266 &  84.97591 &  -2.45913 & II & 20.39 & ... & ... & ... & 15.3 & 13.46 & 12.87 & 12.67 & 12.16 & 11.86 & 11.39 & 10.02 & 5.6 \\ 
SO1268 &  84.97643 &  -2.62190 & TD & $>$23.0 & ... & ... & ... & 17.184 & 14.75 & 14.21 & 13.8 & 13.51 & 13.33 & 13.33 & 13.34 & 9.8 \\ 
SO1269 &  84.97656 &  -2.74664 & III & $<$17 & ... & ... & ... & 11.484 & 10.133 & 9.491 & 9.293 & 9.17 & 9.27 & 9.13 & 9.12 & 9.18 \\ 
SO1274 &  84.97777 &  -2.77618 & II & $<$17 & 15.06 & 14.253 & 13.87 & 12.76 & 11.054 & 10.251 & 9.832 & 9.26 & 8.89 & 8.56 & 7.66 & 5.07 \\ 
SO1275 &  84.97857 &  -2.32504 & III & 17.85 & ... & 15.25 & ... & ... & 12.255 & 11.568 & 11.379 & 11.21 & 11.22 & 11.22 & 11.16 & ... \\ 
SO1282 &  84.98310 &  -2.34357 & III & 18.46 & ... & 15.95 & ... & 13.76 & 12.27 & 11.51 & 11.29 & 11.12 & 11.18 & 11.03 & 11 & ... \\ 
SO1295 &  84.98979 &  -2.53670 & III & 20.78 & 19.2 & 18.22 & 17.6 & 15.17 & 13.31 & 12.69 & 12.36 & 12.1 & 12.03 & 11.97 & 11.92 & ... \\ 
SO1296 &  84.99012 &  -2.79341 & III & 19.27 & ... & 16.34 & ... & ... & 13.227 & 12.529 & 12.339 & 12.21 & 12.24 & 12.21 & 12.15 & ... \\ 
SO1323 &  85.00426 &  -2.33333 & EV & 20.82 & ... & 18.45 & ... & 15.31 & 13.1 & 12.5 & 12.25 & 11.91 & 11.81 & 11.83 & 11.6 & ... \\ 
SO1327 &  85.00816 &  -2.35904 & II & 18.58 & ... & 16.84 & ... & 14.06 & 12.34 & 11.58 & 11.25 & 10.64 & 10.35 & 10.22 & 9.67 & 6.22 \\ 
SO1343 &  85.02133 &  -2.33315 & III & 17.34 & 17 & 14.493 & 14.22 & 12.95 & 11.46 & 10.77 & 10.542 & 10.4 & 10.39 & 10.36 & 10.33 & ... \\ 
SO1347 &  85.02210 &  -2.76069 & III & $<$17 & ... & ... & ... & 11.471 & 10.008 & 9.366 & 9.189 & 9.2 & 9.25 & 9.22 & 9.01 & 9.09 \\ 
SO1353 &  85.02959 &  -2.54577 & III & 20.66 & 19 & 18.15 & 17.9 & 15.28 & 13.42 & 12.81 & 12.54 & 12.23 & 12.15 & 12.19 & 12.1 & ... \\ 
SO1359 &  85.03622 &  -2.54536 & III & 19.47 & ... & 17.16 & ... & 13.9 & 11.77 & 11.15 & 10.85 & 10.59 & 10.51 & 10.48 & 10.48 & ... \\ 
SO1361 &  85.03713 &  -2.55938 & II & $<$17 & 16.2 & 15.1 & ... & 13.907 & 11.501 & 10.546 & 9.911 & 9.1 & 8.72 & 8.33 & 7.5 & ... \\ 
SO1362 &  85.03891 &  -2.41853 & II & 19.8 & ... & 18.27 & ... & 15.15 & 13.15 & 12.5 & 12.15 & 11.51 & 11.09 & 10.66 & 9.87 & ... \\ 

&&&&&&&&&&&&&&&&\\

\hline\hline
\label{phot}
\end{longtable}
\end{landscape}

\clearpage

\end{document}